\newif\ifdoubleblind
\doubleblindfalse

\ifdoubleblind
\documentclass[acmsmall,screen,review,anonymous]{acmart}
\else
\documentclass[acmsmall,screen,nonacm]{acmart}
\fi
\usepackage{microtype}
\usepackage[utf8]{inputenc}
\usepackage[T1]{fontenc}
\usepackage[english]{babel}
\usepackage{csquotes}
\usepackage{enumitem}
\usepackage{booktabs}
\usepackage{tabularx}
\usepackage{tcolorbox}
\usepackage{pgfplots}
\pgfplotsset{compat=1.18}
\usepgfplotslibrary{statistics}
\usepackage{textcomp}
\usepackage{xurl}
\usepackage[htt]{hyphenat}
\usepackage{etoolbox}

\usepackage{tikz}
\usetikzlibrary{positioning, fit, backgrounds, calc, arrows.meta, shapes.geometric, patterns}

\tcbset{
  before skip=3pt,
  after skip=3pt,
  top=1.0mm, bottom=1.0mm,
  left=2.0mm, right=2.0mm,
  boxsep=1pt
}

\title[Diagnosing Generalization Boundaries in AI-Based Vulnerability Detection]{Diagnosing Generalization Boundaries in AI-Based Vulnerability Detection via Calibrated Embedding Distances}

\ifdoubleblind
 \author{Anonymous Author(s)}
 \affiliation{
   \institution{Anonymous Institution}
   \country{Anonymous Country}
 }
\else
\author{Konstantin Ievtushenko}
\email{k\_iev@uni-koblenz.de}
\affiliation{%
  \institution{University of Koblenz}
  \city{Koblenz}
  \country{Germany}
}

\author{Ralf Lämmel}
\email{laemmel@uni-koblenz.de}
\affiliation{%
  \institution{University of Koblenz}
  \city{Koblenz}
  \country{Germany}
}

\author{Jan Jürjens}
\email{juerjens@uni-koblenz.de}
\affiliation{%
  \institution{University of Koblenz}
  \city{Koblenz}
  \country{Germany}
}
\affiliation{%
  \institution{Fraunhofer ISST}
  \city{Dortmund}
  \country{Germany}
}
\fi

\begin{document}

\begin{abstract}
High benchmark scores in learning-based vulnerability detection mask severe degradation under distribution shifts. This paper demonstrates that calibrated embedding geometry exposes these hidden generalization boundaries on syntactically similar vulnerable and patched code pairs. An evaluation of CodeBERT and Qwen2.5-Coder across pattern, project, and temporal shifts on a strictly paired dataset reveals three main findings. First, whitened intermediate-layer embeddings with local nearest-neighbour scoring separate unfamiliar samples with a mean out-of-distribution AUROC of 0.739 (RQ1). Second, distance scores reveal a shift-specific reliability gradient where accuracy drops by over 35 percentage points and cross-project shifts cause the sharpest degradation, even though temporal shifts drift geometrically farthest (RQ2). Third, false positives on patched code systematically correlate with geometric proximity to vulnerable training neighbourhoods, increasing from 9.4\% in the lowest-risk decile to 58.8\% in the highest (RQ3). These calibrated distances establish a diagnostic signal that maps representation geometry directly to model degradation without requiring auxiliary models.
\end{abstract}



\maketitle

\section{Introduction}
\label{sec:introduction}

Alongside traditional static analysis, automated software vulnerability detection increasingly incorporates learning-based techniques \cite{chakraborty2022deep, sheng2025llmssoftwaresecuritysurvey}. In particular, transformer-based code models and Large Language Models (LLMs) show strong benchmark performance, with recent contributions demonstrating the efficacy of fine-tuned code models for security-patch and vulnerability analysis in industrial settings \cite{ran2025efficient}. These models process source code directly and adapt to security-specific classification via fine-tuning \cite{shestov2024finetuning, sheng2025llmssoftwaresecuritysurvey}, suggesting they capture recurring vulnerability patterns that are difficult to express through manual rules.

However, strong benchmark performance does not guarantee reliable generalization beyond the training distribution. Many evaluations rely on in-distribution test sets sharing similar projects, coding styles, time periods, and vulnerability patterns. Under such conditions, models often exploit recurring superficial cues rather than learning robust vulnerability semantics \cite{chakraborty2022deep}, causing detection performance to degrade on unfamiliar projects, later vulnerability instances, or unseen code structures that differ from the training distribution \cite{li2025outofdistribution}.

This raises a central question: where do the generalization boundaries of these models lie? Model predictions are only actionable if there is some indication of whether a model operates in a familiar region of its input space \cite{hendrycks2017baseline, ovadia2019trust}. For security applications, operating outside these boundaries is problematic because both missed vulnerabilities and false alarms carry severe operational costs.

A particularly challenging scenario involves \textit{hard-negative} samples that are difficult to distinguish from positive instances due to high semantic and structural similarity \cite{robinson2021contrastive}. In this study, we operationalize this concept using vulnerable functions and their corresponding patched counterparts as software evolution artifacts. Such pairs share most tokens, control structures, and context, differing only in localized, security-relevant edits \cite{xiao2020mvp}. Unlike arbitrary benign code where models often exploit superficial shortcuts \cite{geirhos2020shortcut}, hard-negative pairs enforce a stricter evaluation of whether the model captures genuine vulnerability semantics.

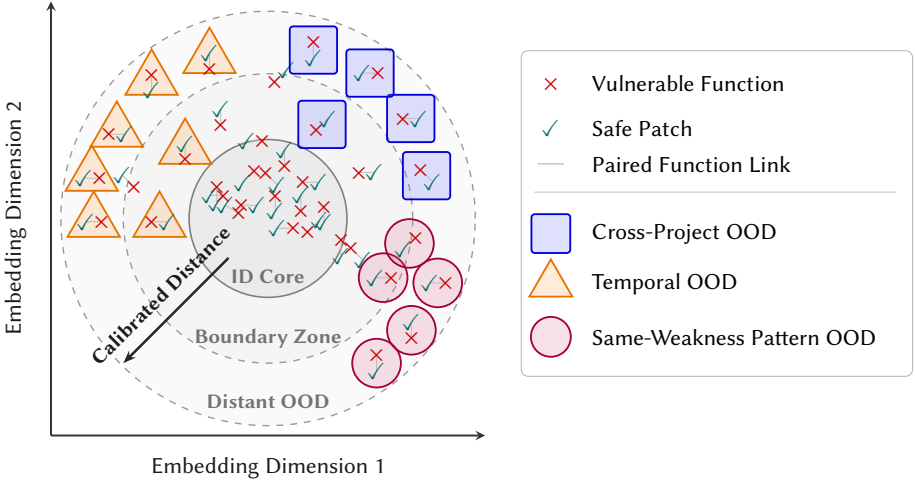
\begin{figure}[!t]
\centering
\resizebox{0.86\linewidth}{!}{%
\begin{tikzpicture}[
    every node/.style={font=\sffamily},
    axis/.style={->, >=stealth, line width=0.85pt, draw=black!80},
    axis_lbl/.style={font=\sffamily\normalsize, text=black!90},
    contour/.style={draw=black!40, dashed, line width=0.6pt},
    zone_txt/.style={font=\sffamily\small\bfseries, text=black!55},
    v_pt/.style={text=red!80!black, font=\normalsize\bfseries, inner sep=0pt},
    p_pt/.style={text=teal!85!black, font=\normalsize\bfseries, inner sep=0pt},
    pair_link/.style={draw=black!25, line width=0.45pt},
    badge_proj/.style={draw=blue!85!black, fill=blue!60, fill opacity=0.20, text opacity=1, line width=0.8pt, rounded corners=1.5pt, minimum size=7.2mm},
    badge_temp/.style={draw=orange!90!black, fill=orange!70, fill opacity=0.22, text opacity=1, line width=0.8pt, regular polygon, regular polygon sides=3, minimum size=9.2mm},
    badge_pat/.style={draw=purple!85!black, fill=purple!60, fill opacity=0.20, text opacity=1, line width=0.8pt, circle, minimum size=7.4mm}
]

    \path[use as bounding box] (0.1, 0.1) rectangle (13.85, 7.35);

    \coordinate (Origin) at (0.65, 0.65);
    \draw[axis] (Origin) -- (7.25, 0.65);
    \node[axis_lbl, anchor=north] at (3.95, 0.45) {Embedding Dimension 1};

    \draw[axis] (Origin) -- (0.65, 7.25);
    \node[axis_lbl, anchor=south, rotate=90] at (0.40, 3.95) {Embedding Dimension 2};

    \coordinate (C) at (3.95, 3.95);

    \draw[contour, fill=black!2] (C) circle (3.15cm);
    \draw[contour, fill=black!4] (C) circle (2.20cm);
    \draw[draw=black!50, fill=black!8, line width=0.7pt] (C) circle (1.20cm);

    \node[zone_txt, anchor=north] at ($(C) + (0, -0.65)$) {ID Core};
    \node[zone_txt, anchor=north] at ($(C) + (0, -1.55)$) {Boundary Zone};
    \node[zone_txt, anchor=north] at ($(C) + (0, -2.55)$) {Distant OOD};

    \draw[pair_link] ($(C) + (0.11, 0.35)$) -- ($(C) + (0.11, 0.11)$);
    \node[v_pt] at ($(C) + (0.11, 0.35)$) {$\times$};
    \node[p_pt] at ($(C) + (0.11, 0.11)$) {$\checkmark$};

    \draw[pair_link] ($(C) + (-0.41, 0.20)$) -- ($(C) + (-0.17, 0.20)$);
    \node[v_pt] at ($(C) + (-0.41, 0.20)$) {$\times$};
    \node[p_pt] at ($(C) + (-0.17, 0.20)$) {$\checkmark$};

    \draw[pair_link] ($(C) + (0.38, -0.15)$) -- ($(C) + (0.14, -0.15)$);
    \node[v_pt] at ($(C) + (0.38, -0.15)$) {$\times$};
    \node[p_pt] at ($(C) + (0.14, -0.15)$) {$\checkmark$};

    \draw[pair_link] ($(C) + (0.50, 0.11)$) -- ($(C) + (0.50, 0.35)$);
    \node[v_pt] at ($(C) + (0.50, 0.11)$) {$\times$};
    \node[p_pt] at ($(C) + (0.50, 0.35)$) {$\checkmark$};

    \draw[pair_link] ($(C) + (-0.04, 0.69)$) -- ($(C) + (0.14, 0.51)$);
    \node[v_pt] at ($(C) + (-0.04, 0.69)$) {$\times$};
    \node[p_pt] at ($(C) + (0.14, 0.51)$) {$\checkmark$};

    \draw[pair_link] ($(C) + (-0.45, 0.08)$) -- ($(C) + (-0.65, 0.20)$);
    \node[v_pt] at ($(C) + (-0.45, 0.08)$) {$\times$};
    \node[p_pt] at ($(C) + (-0.65, 0.20)$) {$\checkmark$};

    \draw[pair_link] ($(C) + (0.60, -0.21)$) -- ($(C) + (0.78, -0.03)$);
    \node[v_pt] at ($(C) + (0.60, -0.21)$) {$\times$};
    \node[p_pt] at ($(C) + (0.78, -0.03)$) {$\checkmark$};

    \draw[pair_link] ($(C) + (-0.21, 0.72)$) -- ($(C) + (-0.39, 0.54)$);
    \node[v_pt] at ($(C) + (-0.21, 0.72)$) {$\times$};
    \node[p_pt] at ($(C) + (-0.39, 0.54)$) {$\checkmark$};

    \draw[pair_link] ($(C) + (0.53, 0.55)$) -- ($(C) + (0.77, 0.55)$);
    \node[v_pt] at ($(C) + (0.53, 0.55)$) {$\times$};
    \node[p_pt] at ($(C) + (0.77, 0.55)$) {$\checkmark$};

    \draw[pair_link] ($(C) + (-0.78, 0.48)$) -- ($(C) + (-0.78, 0.24)$);
    \node[v_pt] at ($(C) + (-0.78, 0.48)$) {$\times$};
    \node[p_pt] at ($(C) + (-0.78, 0.24)$) {$\checkmark$};

    \draw[pair_link] ($(C) + (0.85, 0.18)$) -- ($(C) + (0.85, -0.06)$);
    \node[v_pt] at ($(C) + (0.85, 0.18)$) {$\times$};
    \node[p_pt] at ($(C) + (0.85, -0.06)$) {$\checkmark$};

    \draw[pair_link] ($(C) + (0.25, 0.80)$) -- ($(C) + (0.25, 1.04)$);
    \node[v_pt] at ($(C) + (0.25, 0.80)$) {$\times$};
    \node[p_pt] at ($(C) + (0.25, 1.04)$) {$\checkmark$};

    \draw[pair_link] ($(C) + (-0.68, 0.35)$) -- ($(C) + (-0.88, 0.35)$);
    \node[v_pt] at ($(C) + (-0.68, 0.35)$) {$\times$};
    \node[p_pt] at ($(C) + (-0.88, 0.35)$) {$\checkmark$};

    \draw[pair_link] ($(C) + (-0.09, 1.18)$) -- ($(C) + (-0.33, 1.18)$);
    \node[v_pt] at ($(C) + (-0.09, 1.18)$) {$\times$};
    \node[p_pt] at ($(C) + (-0.33, 1.18)$) {$\checkmark$};

    \draw[pair_link] ($(C) + (1.11, -0.33)$) -- ($(C) + (1.11, -0.57)$);
    \node[v_pt] at ($(C) + (1.11, -0.33)$) {$\times$};
    \node[p_pt] at ($(C) + (1.11, -0.57)$) {$\checkmark$};

    \draw[pair_link] ($(C) + (1.38, 0.70)$) -- ($(C) + (1.62, 0.70)$);
    \node[v_pt] at ($(C) + (1.38, 0.70)$) {$\times$};
    \node[p_pt] at ($(C) + (1.62, 0.70)$) {$\checkmark$};

    \draw[pair_link] ($(C) + (-0.72, 1.42)$) -- ($(C) + (-0.72, 1.66)$);
    \node[v_pt] at ($(C) + (-0.72, 1.42)$) {$\times$};
    \node[p_pt] at ($(C) + (-0.72, 1.66)$) {$\checkmark$};

    \draw[pair_link] ($(C) + (1.26, -0.45)$) -- ($(C) + (1.44, -0.63)$);
    \node[v_pt] at ($(C) + (1.26, -0.45)$) {$\times$};
    \node[p_pt] at ($(C) + (1.44, -0.63)$) {$\checkmark$};

    \draw[pair_link] ($(C) + (-2.04, 0.48)$) -- ($(C) + (-2.22, 0.66)$);
    \node[v_pt] at ($(C) + (-2.04, 0.48)$) {$\times$};
    \node[p_pt] at ($(C) + (-2.22, 0.66)$) {$\checkmark$};

    \draw[pair_link] ($(C) + (0.10, 2.08)$) -- ($(C) + (0.28, 2.26)$);
    \node[v_pt] at ($(C) + (0.10, 2.08)$) {$\times$};
    \node[p_pt] at ($(C) + (0.28, 2.26)$) {$\checkmark$};

    \node[badge_proj] at ($(C) + (0.82, 1.43)$) {};
    \draw[pair_link] ($(C) + (0.73, 1.34)$) -- ($(C) + (0.91, 1.52)$);
    \node[v_pt] at ($(C) + (0.73, 1.34)$) {$\times$};
    \node[p_pt] at ($(C) + (0.91, 1.52)$) {$\checkmark$};

    \node[badge_temp] at ($(C) + (-1.26, 1.06)$) {};
    \draw[pair_link] ($(C) + (-1.26, 0.90)$) -- ($(C) + (-1.26, 1.14)$);
    \node[v_pt] at ($(C) + (-1.26, 0.90)$) {$\times$};
    \node[p_pt] at ($(C) + (-1.26, 1.14)$) {$\checkmark$};

    \node[badge_temp] at ($(C) + (-1.65, -0.05)$) {};
    \draw[pair_link] ($(C) + (-1.77, -0.05)$) -- ($(C) + (-1.53, -0.05)$);
    \node[v_pt] at ($(C) + (-1.77, -0.05)$) {$\times$};
    \node[p_pt] at ($(C) + (-1.53, -0.05)$) {$\checkmark$};

    \node[badge_pat] at ($(C) + (1.75, -0.90)$) {};
    \draw[pair_link] ($(C) + (1.87, -0.90)$) -- ($(C) + (1.63, -0.90)$);
    \node[v_pt] at ($(C) + (1.87, -0.90)$) {$\times$};
    \node[p_pt] at ($(C) + (1.63, -0.90)$) {$\checkmark$};

    \node[badge_pat] at ($(C) + (2.15, -0.38)$) {};
    \draw[pair_link] ($(C) + (2.24, -0.29)$) -- ($(C) + (2.06, -0.47)$);
    \node[v_pt] at ($(C) + (2.24, -0.29)$) {$\times$};
    \node[p_pt] at ($(C) + (2.06, -0.47)$) {$\checkmark$};

    \node[badge_proj] at ($(C) + (2.41, 0.65)$) {};
    \draw[pair_link] ($(C) + (2.32, 0.74)$) -- ($(C) + (2.50, 0.56)$);
    \node[v_pt] at ($(C) + (2.32, 0.74)$) {$\times$};
    \node[p_pt] at ($(C) + (2.50, 0.56)$) {$\checkmark$};

    \node[badge_proj] at ($(C) + (2.17, 1.52)$) {};
    \draw[pair_link] ($(C) + (2.05, 1.52)$) -- ($(C) + (2.29, 1.52)$);
    \node[v_pt] at ($(C) + (2.05, 1.52)$) {$\times$};
    \node[p_pt] at ($(C) + (2.29, 1.52)$) {$\checkmark$};

    \node[badge_proj] at ($(C) + (1.55, 2.21)$) {};
    \draw[pair_link] ($(C) + (1.67, 2.21)$) -- ($(C) + (1.43, 2.21)$);
    \node[v_pt] at ($(C) + (1.67, 2.21)$) {$\times$};
    \node[p_pt] at ($(C) + (1.43, 2.21)$) {$\checkmark$};

    \node[badge_proj] at ($(C) + (0.69, 2.56)$) {};
    \draw[pair_link] ($(C) + (0.69, 2.68)$) -- ($(C) + (0.69, 2.44)$);
    \node[v_pt] at ($(C) + (0.69, 2.68)$) {$\times$};
    \node[p_pt] at ($(C) + (0.69, 2.44)$) {$\checkmark$};

    \node[badge_temp] at ($(C) + (-0.89, 2.44)$) {};
    \draw[pair_link] ($(C) + (-0.89, 2.28)$) -- ($(C) + (-0.89, 2.52)$);
    \node[v_pt] at ($(C) + (-0.89, 2.28)$) {$\times$};
    \node[p_pt] at ($(C) + (-0.89, 2.52)$) {$\checkmark$};

    \node[badge_temp] at ($(C) + (-1.77, 2.11)$) {};
    \draw[pair_link] ($(C) + (-1.77, 2.19)$) -- ($(C) + (-1.77, 1.95)$);
    \node[v_pt] at ($(C) + (-1.77, 2.19)$) {$\times$};
    \node[p_pt] at ($(C) + (-1.77, 1.95)$) {$\checkmark$};

    \node[badge_temp] at ($(C) + (-2.30, 1.32)$) {};
    \draw[pair_link] ($(C) + (-2.42, 1.28)$) -- ($(C) + (-2.18, 1.28)$);
    \node[v_pt] at ($(C) + (-2.42, 1.28)$) {$\times$};
    \node[p_pt] at ($(C) + (-2.18, 1.28)$) {$\checkmark$};

    \node[badge_temp] at ($(C) + (-2.68, 0.65)$) {};
    \draw[pair_link] ($(C) + (-2.56, 0.61)$) -- ($(C) + (-2.80, 0.61)$);
    \node[v_pt] at ($(C) + (-2.56, 0.61)$) {$\times$};
    \node[p_pt] at ($(C) + (-2.80, 0.61)$) {$\checkmark$};

    \node[badge_temp] at ($(C) + (-2.65, -0.05)$) {};
    \draw[pair_link] ($(C) + (-2.53, -0.05)$) -- ($(C) + (-2.77, -0.05)$);
    \node[v_pt] at ($(C) + (-2.53, -0.05)$) {$\times$};
    \node[p_pt] at ($(C) + (-2.77, -0.05)$) {$\checkmark$};

    \node[badge_pat] at ($(C) + (1.65, -2.20)$) {};
    \draw[pair_link] ($(C) + (1.65, -2.08)$) -- ($(C) + (1.65, -2.32)$);
    \node[v_pt] at ($(C) + (1.65, -2.08)$) {$\times$};
    \node[p_pt] at ($(C) + (1.65, -2.32)$) {$\checkmark$};

    \node[badge_pat] at ($(C) + (2.18, -1.70)$) {};
    \draw[pair_link] ($(C) + (2.18, -1.82)$) -- ($(C) + (2.18, -1.58)$);
    \node[v_pt] at ($(C) + (2.18, -1.82)$) {$\times$};
    \node[p_pt] at ($(C) + (2.18, -1.58)$) {$\checkmark$};

    \node[badge_pat] at ($(C) + (2.58, -0.98)$) {};
    \draw[pair_link] ($(C) + (2.70, -0.98)$) -- ($(C) + (2.46, -0.98)$);
    \node[v_pt] at ($(C) + (2.70, -0.98)$) {$\times$};
    \node[p_pt] at ($(C) + (2.46, -0.98)$) {$\checkmark$};

    \draw[-{Stealth[length=5.5pt]}, line width=1.1pt, draw=black!85] 
        ($(C) + (-0.60, -0.60)$) -- ($(C) + (-2.20, -2.20)$)
        node[midway, sloped, above=2.5pt, font=\sffamily\small\bfseries, text=black!90] 
        {Calibrated Distance};

    \coordinate (L) at (7.80, 6.50);
    \draw[draw=black!30, fill=white, rounded corners=3pt] (7.80, 1.55) rectangle (13.75, 6.50);

    \node[v_pt] at ($(L) + (0.45, -0.50)$) {$\times$};
    \node[anchor=west, font=\sffamily\normalsize] at ($(L) + (0.95, -0.50)$) {Vulnerable Function};

    \node[p_pt] at ($(L) + (0.45, -1.18)$) {$\checkmark$};
    \node[anchor=west, font=\sffamily\normalsize] at ($(L) + (0.95, -1.18)$) {Safe Patch};

    \draw[pair_link, line width=0.6pt] ($(L) + (0.25, -1.72)$) -- ($(L) + (0.65, -1.72)$);
    \node[anchor=west, font=\sffamily\normalsize] at ($(L) + (0.95, -1.72)$) {Paired Function Link};
    \draw[draw=black!20, line width=0.5pt] ($(L) + (0.20, -2.15)$) -- ($(L) + (5.70, -2.15)$);

    \node[badge_proj, scale=0.82] at ($(L) + (0.45, -2.78)$) {};
    \node[anchor=west, font=\sffamily\normalsize] at ($(L) + (0.95, -2.78)$) {Cross-Project OOD};

    \node[badge_temp, scale=0.82] at ($(L) + (0.45, -3.55)$) {};
    \node[anchor=west, font=\sffamily\normalsize] at ($(L) + (0.95, -3.55)$) {Temporal OOD};

    \node[badge_pat, scale=0.82] at ($(L) + (0.45, -4.35)$) {};
    \node[anchor=west, font=\sffamily\normalsize] at ($(L) + (0.95, -4.35)$) {Same-Weakness Pattern OOD};

\end{tikzpicture}%
}
\caption{Conceptual 2D representation space motivating this study. In-distribution code forms a dense core of paired functions ($\times, \checkmark$) exhibiting high syntactic overlap despite opposite labels. Under distribution shift, samples drift outward across boundary and distant zones, where calibrated embedding distance exposes model generalization limits.}
\Description{Conceptual 2D scatter plot illustrating how in-distribution pairs form a dense core, while distribution shifts drift outward into boundary and distant zones.}
\label{fig:motivation_geometry}
\end{figure}

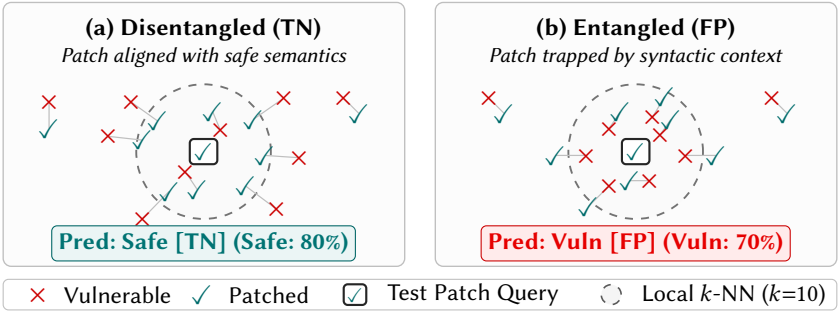
\begin{figure}[t]
\centering
\resizebox{0.80\linewidth}{!}{%
\begin{tikzpicture}[
    every node/.style={font=\sffamily},
    >=stealth,
    v_pt/.style={text=red!80!black, font=\sffamily\normalsize\bfseries, inner sep=0pt},
    p_pt/.style={text=teal!85!black, font=\sffamily\normalsize\bfseries, inner sep=0pt},
    pair_link/.style={draw=black!25, line width=0.45pt},
    query/.style={draw=black!85, fill=white, line width=0.8pt, rounded corners=1.5pt, inner sep=1.5pt},
    kcircle/.style={draw=black!55, dashed, line width=0.7pt, fill=black!3, fill opacity=0.40},
    panel/.style={draw=black!25, fill=gray!2, rounded corners=3pt, line width=0.5pt},
    badge_tn/.style={draw=teal!80!black, fill=teal!8, text=teal!90!black, rounded corners=2pt, font=\sffamily\small\bfseries, inner sep=2.5pt},
    badge_fp/.style={draw=red!80!black, fill=red!8, text=red!90!black, rounded corners=2pt, font=\sffamily\small\bfseries, inner sep=2.5pt}
]

    \path[use as bounding box] (0, -0.15) rectangle (11.00, 4.20);

    \begin{scope}[shift={(0,0)}]
        \draw[panel] (0.05, 0.65) rectangle (5.30, 4.10);
        \node[anchor=north, align=center] at (2.67, 4.02) {\textbf{\small (a) Disentangled (TN)}\\[-2pt]\textit{\footnotesize Patch aligned with safe semantics}};

        \coordinate (Q1) at (2.67, 2.15);
        \draw[kcircle] (Q1) circle (0.88cm);
        \node[query] at (Q1) {\textcolor{teal!85!black}{\small\bfseries $\checkmark$}};

        \draw[pair_link] ($(Q1) + (0.22, 0.28)$) -- ($(Q1) + (0.12, 0.52)$);
        \node[v_pt] at ($(Q1) + (0.22, 0.28)$) {$\times$};
        \node[p_pt] at ($(Q1) + (0.12, 0.52)$) {$\checkmark$};

        \draw[pair_link] ($(Q1) + (-0.24, -0.26)$) -- ($(Q1) + (-0.08, -0.50)$);
        \node[v_pt] at ($(Q1) + (-0.24, -0.26)$) {$\times$};
        \node[p_pt] at ($(Q1) + (-0.08, -0.50)$) {$\checkmark$};

        \draw[pair_link] ($(Q1) + (-0.95, 0.65)$) -- ($(Q1) + (-0.60, 0.40)$);
        \node[v_pt] at ($(Q1) + (-0.95, 0.65)$) {$\times$};
        \node[p_pt] at ($(Q1) + (-0.60, 0.40)$) {$\checkmark$};

        \draw[pair_link] ($(Q1) + (-1.25, 0.20)$) -- ($(Q1) + (-0.80, 0.15)$);
        \node[v_pt] at ($(Q1) + (-1.25, 0.20)$) {$\times$};
        \node[p_pt] at ($(Q1) + (-0.80, 0.15)$) {$\checkmark$};

        \draw[pair_link] ($(Q1) + (1.05, 0.70)$) -- ($(Q1) + (0.65, 0.42)$);
        \node[v_pt] at ($(Q1) + (1.05, 0.70)$) {$\times$};
        \node[p_pt] at ($(Q1) + (0.65, 0.42)$) {$\checkmark$};

        \draw[pair_link] ($(Q1) + (1.25, -0.08)$) -- ($(Q1) + (0.80, -0.05)$);
        \node[v_pt] at ($(Q1) + (1.25, -0.08)$) {$\times$};
        \node[p_pt] at ($(Q1) + (0.80, -0.05)$) {$\checkmark$};

        \draw[pair_link] ($(Q1) + (0.95, -0.75)$) -- ($(Q1) + (0.55, -0.45)$);
        \node[v_pt] at ($(Q1) + (0.95, -0.75)$) {$\times$};
        \node[p_pt] at ($(Q1) + (0.55, -0.45)$) {$\checkmark$};

        \draw[pair_link] ($(Q1) + (-0.80, -0.88)$) -- ($(Q1) + (-0.45, -0.52)$);
        \node[v_pt] at ($(Q1) + (-0.80, -0.88)$) {$\times$};
        \node[p_pt] at ($(Q1) + (-0.45, -0.52)$) {$\checkmark$};

        \draw[pair_link] (0.65, 2.80) -- (0.65, 2.45);
        \node[v_pt] at (0.65, 2.80) {$\times$};
        \node[p_pt] at (0.65, 2.45) {$\checkmark$};

        \draw[pair_link] (4.50, 2.85) -- (4.70, 2.65);
        \node[v_pt] at (4.50, 2.85) {$\times$};
        \node[p_pt] at (4.70, 2.65) {$\checkmark$};

        \node[badge_tn] at (2.67, 0.95) {Pred: Safe [TN] (Safe: 80\%)};
    \end{scope}

    \begin{scope}[shift={(5.65,0)}]
        \draw[panel] (0.05, 0.65) rectangle (5.30, 4.10);
        \node[anchor=north, align=center] at (2.67, 4.02) {\textbf{\small (b) Entangled (FP)}\\[-2pt]\textit{\footnotesize Patch trapped by syntactic context}};

        \coordinate (Q2) at (2.67, 2.15);
        \draw[kcircle] (Q2) circle (0.88cm);
        \node[query] at (Q2) {\textcolor{teal!85!black}{\small\bfseries $\checkmark$}};

        \draw[pair_link] ($(Q2) + (-0.35, 0.30)$) -- ($(Q2) + (-0.20, 0.52)$);
        \node[v_pt] at ($(Q2) + (-0.35, 0.30)$) {$\times$};
        \node[p_pt] at ($(Q2) + (-0.20, 0.52)$) {$\checkmark$};

        \draw[pair_link] ($(Q2) + (0.32, 0.22)$) -- ($(Q2) + (0.46, 0.42)$);
        \node[v_pt] at ($(Q2) + (0.32, 0.22)$) {$\times$};
        \node[p_pt] at ($(Q2) + (0.46, 0.42)$) {$\checkmark$};

        \draw[pair_link] ($(Q2) + (0.18, -0.38)$) -- ($(Q2) + (-0.08, -0.38)$);
        \node[v_pt] at ($(Q2) + (0.18, -0.38)$) {$\times$};
        \node[p_pt] at ($(Q2) + (-0.08, -0.38)$) {$\checkmark$};

        \draw[pair_link] ($(Q2) + (0.22, 0.45)$) -- ($(Q2) + (0.38, 0.72)$);
        \node[v_pt] at ($(Q2) + (0.22, 0.45)$) {$\times$};
        \node[p_pt] at ($(Q2) + (0.38, 0.72)$) {$\checkmark$};

        \draw[pair_link] ($(Q2) + (-0.65, -0.05)$) -- ($(Q2) + (-1.05, -0.05)$);
        \node[v_pt] at ($(Q2) + (-0.65, -0.05)$) {$\times$};
        \node[p_pt] at ($(Q2) + (-1.05, -0.05)$) {$\checkmark$};

        \draw[pair_link] ($(Q2) + (0.65, -0.05)$) -- ($(Q2) + (1.05, -0.05)$);
        \node[v_pt] at ($(Q2) + (0.65, -0.05)$) {$\times$};
        \node[p_pt] at ($(Q2) + (1.05, -0.05)$) {$\checkmark$};

        \draw[pair_link] ($(Q2) + (-0.35, -0.45)$) -- ($(Q2) + (-0.62, -0.72)$);
        \node[v_pt] at ($(Q2) + (-0.35, -0.45)$) {$\times$};
        \node[p_pt] at ($(Q2) + (-0.62, -0.72)$) {$\checkmark$};

        \draw[pair_link] (0.75, 2.85) -- (0.95, 2.65);
        \node[v_pt] at (0.75, 2.85) {$\times$};
        \node[p_pt] at (0.95, 2.65) {$\checkmark$};

        \draw[pair_link] (4.45, 2.85) -- (4.65, 2.65);
        \node[v_pt] at (4.45, 2.85) {$\times$};
        \node[p_pt] at (4.65, 2.65) {$\checkmark$};

        \node[badge_fp] at (2.67, 0.95) {Pred: Vuln [FP] (Vuln: 70\%)};
    \end{scope}

    \draw[draw=black!25, fill=white, rounded corners=2pt] (0.05, 0.08) rectangle (10.95, 0.48);

    \node[v_pt] at (0.50, 0.28) {$\times$};
    \node[anchor=west, font=\sffamily\small] at (0.72, 0.28) {Vulnerable};

    \node[p_pt] at (2.65, 0.28) {$\checkmark$};
    \node[anchor=west, font=\sffamily\small] at (2.87, 0.28) {Patched};

    \node[query, scale=0.82] at (4.65, 0.28) {\textcolor{teal!85!black}{\small\bfseries $\checkmark$}};
    \node[anchor=west, font=\sffamily\small] at (4.95, 0.28) {Test Patch Query};

    \draw[draw=black!65, dashed, line width=0.6pt, fill=black!4] (8.00, 0.28) circle (0.13cm);
    \node[anchor=west, font=\sffamily\small] at (8.27, 0.28) {Local $k$-NN ($k{=}10$)};

\end{tikzpicture}%
}
\caption{Local neighbourhood proximity mechanism. Although training data is paired, latent distances are asymmetric: (a) A patch whose top-$k$ neighbours are dominated by safe samples yields a True Negative. (b) Syntactic overlap traps a patch closer to vulnerable counterparts, inducing a False Positive.}
\Description{Diagram comparing two local k-NN neighbourhoods: panel (a) shows a patch query surrounded by safe samples resulting in a true negative, while panel (b) shows syntactic overlap trapping the patch within vulnerable training samples, inducing a false positive.}
\label{fig:rq3_motivation_neighbourhood}
\end{figure}

Ideally, an embedding space should organize code fragments by security semantics. In practice, however, vulnerable and patched snippets often cluster closely due to syntactic overlap, while out-of-distribution (OOD) samples drift far from training neighbourhoods (Figure~\ref{fig:motivation_geometry}). Moreover, at the local level this geometric overlap can trap samples within opposing neighbourhoods, most critically pulling safe patches into vulnerable clusters to trigger false alarms (Figure~\ref{fig:rq3_motivation_neighbourhood}). Standard aggregate metrics reflect overall correctness, but fail to diagnose either phenomenon.

This paper investigates generalization boundaries through embedding-based distance metrics, hypothesizing that geometric proximity to the training distribution provides a diagnostic reliability signal \cite{papernot2018deepknn, sun2022knnood}. While contextual transformer representations generally suffer from inherent anisotropy \cite{ethayarajh2019contextual}, local neighbourhood structures can capture vulnerability similarity in code \cite{shimmi2024vulsim}. Yet, whether calibrated embedding geometry reliably exposes model failure boundaries under hard-negative and OOD shifts has remained an open challenge that we resolve in this work. To systematically examine this, an encoder (CodeBERT) and a generative LLM (Qwen2.5-Coder) are evaluated across controlled shift scenarios guided by three research questions:

\begin{itemize}
\item \textbf{RQ1 (Embedding Space Calibration):} \textit{Which representation choices improve the separability of in-distribution and out-of-distribution samples in the embedding space?}
\item \textbf{RQ2 (Generalization Boundaries):} \textit{To what extent do embedding-based distance metrics reveal generalization boundaries, and how do these boundaries behave under specific distribution shifts?}
\item \textbf{RQ3 (Vulnerable-Neighbourhood Proximity):} \textit{To what extent are false positives on patched hard-negatives characterized by geometric proximity to vulnerable training neighbourhoods?}
\end{itemize}

By addressing these questions, this paper makes three main contributions:

\begin{itemize} 
    \item \textbf{A paired hard-negative ID/OOD benchmark.} We construct an open, leakage-free dataset of strictly paired vulnerable and patched C/C++ functions partitioned across controlled pattern, project, and temporal shifts, preventing models from exploiting superficial shortcuts. 
    \item \textbf{Distance-based generalization boundaries.} We demonstrate that model degradation correlates directly with representation distance, showing that calibrated intermediate embeddings expose a continuous reliability gradient where accuracy drops by over 35 percentage points toward chance level under distribution shift.
    \item \textbf{Geometric diagnostics and test-time triage.} We identify local neighbourhood entrapment as a primary driver of false alarms on repaired code and demonstrate that proximity to vulnerable training clusters provides an actionable signal to flag uncertain alerts before reaching developers.
\end{itemize}

\section{Dataset and Experimental Design}
\label{sec:dataset_experimental_design}

This section outlines the empirical foundation and experimental design. To analyze generalization boundaries effectively, the dataset must provide reliable vulnerability--patch relations, detailed project metadata, and controlled conditions for distribution shift evaluation.

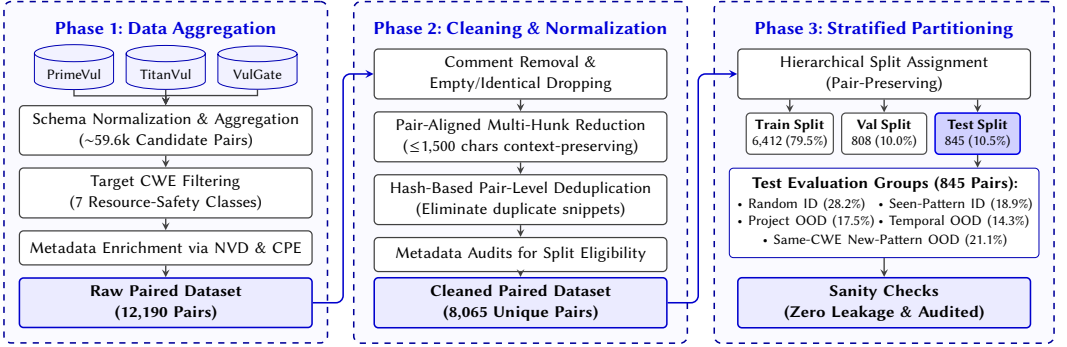
\begin{figure*}[t]
\centering
\resizebox{\linewidth}{!}{%
\begin{tikzpicture}[
    node distance=2.2mm,
    every node/.style={font=\sffamily\footnotesize, text=black},
    card/.style={draw=blue!60!black, dashed, line width=0.75pt, fill=blue!2, rounded corners=3.5pt, inner sep=2.5mm},
    header/.style={font=\sffamily\small\bfseries, align=center, text=blue!85!black, inner sep=0.8mm},
    proc/.style={draw=black!75, line width=0.6pt, fill=white, align=center, rounded corners=2pt, text width=4.75cm, minimum height=0.58cm, inner sep=1.2mm, font=\sffamily\footnotesize},
    artifact/.style={draw=blue!80!black, line width=0.75pt, fill=blue!8, align=center, rounded corners=2pt, text width=4.75cm, minimum height=0.64cm, inner sep=1.2mm, font=\sffamily\footnotesize\bfseries},
    db/.style={shape=cylinder, draw=blue!70!black, shape border rotate=90, aspect=0.25, align=center, fill=blue!5, minimum height=0.45cm, minimum width=1.35cm, font=\sffamily\scriptsize},
    splitnode/.style={draw=black!70, line width=0.6pt, fill=white, align=center, rounded corners=2pt, inner sep=1.0mm, font=\sffamily\scriptsize},
    arr/.style={->, >=stealth, line width=0.65pt, draw=black!75},
    bridge/.style={->, >=stealth, line width=0.75pt, draw=blue!80!black, rounded corners=3.5pt}
]

    \coordinate (c1) at (0, 0);

    \node[header] (h1) at (c1) {Phase 1: Data Aggregation};

    \node[db] (db2) at ($(c1) + (0, -0.75cm)$) {TitanVul};
    \node[db, left=2.0mm of db2] (db1) {PrimeVul};
    \node[db, right=2.0mm of db2] (db3) {VulGate};

    \node[proc, below=2.6mm of db2] (p1_norm) {Schema Normalization \& Aggregation\\($\sim$59.6k Candidate Pairs)};
    \node[proc, below=2.0mm of p1_norm] (p1_cwe) {Target CWE Filtering\\(7 Resource-Safety Classes)};
    \node[proc, below=2.0mm of p1_cwe] (p1_nvd) {Metadata Enrichment via NVD \& CPE};
    \node[artifact, below=2.0mm of p1_nvd] (p1_out) {Raw Paired Dataset\\(12,190 Pairs)};

    \draw[arr] (db1.south) -- ++(0,-0.10) -| (p1_norm.north);
    \draw[arr] (db2.south) -- (p1_norm.north);
    \draw[arr] (db3.south) -- ++(0,-0.10) -| (p1_norm.north);
    \draw[arr] (p1_norm) -- (p1_cwe);
    \draw[arr] (p1_cwe) -- (p1_nvd);
    \draw[arr] (p1_nvd) -- (p1_out);

    \coordinate (c2) at (6.05cm, 0);

    \node[header] (h2) at (c2) {Phase 2: Cleaning \& Normalization};

    \node[proc] (p2_clean) at ($(c2) + (0, -0.75cm)$) {Comment Removal \& Empty/Identical Dropping};
    \node[proc, below=2.0mm of p2_clean] (p2_hunk) {Pair-Aligned Multi-Hunk Reduction\\($\le$1,500 chars context-preserving)};
    \node[proc, below=2.0mm of p2_hunk] (p2_dedup) {Hash-Based Pair-Level Deduplication\\(Eliminate duplicate snippets)};
    \node[proc, below=2.0mm of p2_dedup] (p2_audit) {Metadata Audits for Split Eligibility};
    \node[artifact, at=(p1_out -| c2)] (p2_out) {Cleaned Paired Dataset\\(8,065 Unique Pairs)};

    \draw[arr] (p2_clean) -- (p2_hunk);
    \draw[arr] (p2_hunk) -- (p2_dedup);
    \draw[arr] (p2_dedup) -- (p2_audit);
    \draw[arr] (p2_audit) -- (p2_out);

    \coordinate (c3) at (12.25cm, 0);

    \node[header] (h3) at (c3) {Phase 3: Stratified Partitioning};

    \node[proc] (p3_split) at ($(c3) + (0, -0.75cm)$) {Hierarchical Split Assignment\\(Pair-Preserving)};

    \node[splitnode, minimum width=1.45cm, below=2.2mm of p3_split] (p3_val) {\textbf{Val Split}\\808 (10.0\%)};
    \node[splitnode, minimum width=1.48cm, left=1.2mm of p3_val] (p3_train) {\textbf{Train Split}\\6,412 (79.5\%)};
    \node[splitnode, minimum width=1.45cm, right=1.2mm of p3_val, fill=blue!18, draw=blue!90!black, line width=0.75pt] (p3_test) {\textbf{Test Split}\\845 (10.5\%)};

    \node[draw=blue!70!black, fill=white, rounded corners=2pt, below=2.2mm of p3_val, inner sep=1.4mm, text width=5cm] (p3_groups) {
        \centering
        \textbf{Test Evaluation Groups (845 Pairs):}\\[1.5pt]
        \scriptsize
        \setlength{\tabcolsep}{1.0pt}
        \begin{tabular}{@{}ll@{}}
            \textbullet~Random ID (28.2\%) & \textbullet~Seen-Pattern ID (18.9\%) \\
            \textbullet~Project OOD (17.5\%) & \textbullet~Temporal OOD (14.3\%) \\[1.5pt]
            \multicolumn{2}{@{}c@{}}{\textbullet~Same-CWE New-Pattern OOD (21.1\%)}
        \end{tabular}
    };

    \node[artifact, at=(p1_out -| c3)] (p3_sanity) {Sanity Checks\\(Zero Leakage \& Audited)};

    \draw[arr] (p3_split.south -| p3_train.north) -- (p3_train.north);
    \draw[arr] (p3_split.south -| p3_val.north) -- (p3_val.north);
    \draw[arr] (p3_split.south -| p3_test.north) -- (p3_test.north);

    \draw[arr] (p3_test.south) -- (p3_test.south |- p3_groups.north);
    \draw[arr] (p3_groups.south) -- (p3_sanity.north);

    \coordinate (bridge12) at ($(c1)!0.5!(c2)$);
    \draw[bridge] (p1_out.east) -- (p1_out.east -| bridge12) |- (p2_clean.west);

    \coordinate (bridge23) at (9.08cm, 0);
    \draw[bridge] (p2_out.east) -- (p2_out.east -| bridge23) |- (p3_split.west);

    \begin{scope}[on background layer]
        \node[card, fit=(h1) (db1) (db3) (p1_out)] (card1) {};
        \node[card, fit=(h2) (p2_clean) (p2_out)] (card2) {};
        \node[card, fit=(h3) (p3_split) (p3_groups) (p3_sanity)] (card3) {};
    \end{scope}

\end{tikzpicture}%
}
\caption{Dataset construction pipeline: from multi-source aggregation (Phase 1), through context-preserving cleaning and deduplication (Phase 2), to leakage-free hierarchical partitioning into in-distribution and out-of-distribution groups (Phase 3).}
\Description{Flowchart depicting the three phases of dataset construction: Phase 1 multi-source aggregation from PrimeVul, TitanVul, and VulGate; Phase 2 cleaning and context-preserving deduplication yielding 8,065 unique pairs; and Phase 3 stratified partitioning into 5 distinct evaluation groups.}
\label{fig_dataset_construction_pipeline}
\end{figure*}

Mining vulnerability data directly from repositories introduces risks such as label noise, non-atomic commits, duplicate samples, and incomplete mappings \cite{croft2023dataquality}. Such artifacts can lead to overly optimistic evaluations, causing models to learn dataset-specific shortcuts rather than vulnerability-relevant semantics \cite{chakraborty2022deep}. To prevent this, the present study relies on a curated data foundation combining modern vulnerability datasets with explicit pair information. Figure~\ref{fig_dataset_construction_pipeline} illustrates the three-stage construction pipeline.

\subsection{Data Foundation and Hard-Negative Design}
\label{subsec:data_foundation_aggregation}

The dataset is constructed from three complementary sources: PrimeVul \cite{ding2024vulnerability}, TitanVul \cite{li2025outofdistribution}, and VulGate \cite{vulgate}. This multi-source foundation provides curated labels, explicit vulnerability--patch relations, and the metadata required for controlled distribution shift evaluation. To prevent the representation space from becoming sparse, the dataset is restricted to a semantically coherent subset of memory- and resource-safety vulnerabilities based on their Common Weakness Enumeration (CWE) classes (e.g., CWE-119, CWE-416, CWE-787). Where project information is incomplete, metadata enrichment is performed using the National Vulnerability Database \cite{nvdapi}.

By design, the dataset is constructed exclusively from matched vulnerable and patched functions, yielding a balanced foundation of 12,190 initial pairs. We deliberately exclude arbitrary benign code. Unrelated benign samples introduce distributional disparities that encourage shortcut learning \cite{geirhos2020shortcut}, where models exploit superficial cues such as project-specific identifiers, coding styles, or function complexity rather than genuine flaws \cite{chakraborty2022deep}. Evaluating strictly against patched counterparts forces models to distinguish the localized security edit rather than relying on repository artifacts~\cite{robinson2021contrastive, xiao2020mvp}, aligning with paired evaluation paradigms recently explored in vulnerability benchmarking across code patches and agent protocols~\cite{tang2026benchmarking}.

\subsection{Data Cleaning and Normalization}
\label{subsec:cleaning_length_deduplication}

Transformer-based models typically process a fixed maximum token sequence. Simple truncation of long functions can inadvertently remove the vulnerability's root cause from the model input \cite{rahman2023slicing}. To address this, a character-level safety threshold of 1,500 characters is applied. If a pair exceeds this limit, pair-aligned multi-hunk extraction isolates the line-level differences alongside up to eight lines of surrounding context. Keeping all changed regions prevents the loss of disjoint security edits \cite{nashid2025multihunk} and aligns with patch analysis principles, which emphasize that capturing localized security modifications alongside their immediate syntactic context is essential to preserve semantic consistency across disjoint edits \cite{pan2026mitigating}. While real-world deployment requires scanning full functions without advance knowledge of patches, evaluating full functions directly under a 512-token limit risks discarding the vulnerability root cause entirely \cite{rahman2023slicing}. Pair-aligned hunk reduction accepts a controlled granularity shift in order to guarantee that models are evaluated strictly on their ability to recognize the localized security delta, rather than being penalized by arbitrary tokenizer clipping artifacts.

Following the preprocessing protocol of MCL-VD \cite{mclvd}, inline and block comments are stripped, and pairs that become empty or identical after cleaning are removed. Subsequently, hash-based deduplication is applied at the pair level. If either side of a pair duplicates an existing snippet, the entire pair is discarded. This prevents data leakage and ensures models are evaluated on genuinely unseen samples \cite{allamanis2019adverse, wu2024condefects}. Finally, split-eligibility audits verify that all retained pairs have complete project, temporal, and diff-cluster metadata required for partition assignment, yielding a final dataset of 8,065 verified pairs. To guarantee absolute comparability, both the CodeBERT encoder and the Qwen2.5-Coder LLM are trained and evaluated on this exact same set of normalized inputs.

\subsection{Controlled Evaluation Groups}

Conventional random splits often overestimate generalization because training and test sets share projects, time periods, and recurring fix patterns. To isolate the exact causes of model failure, the dataset is partitioned hierarchically into five distinct evaluation groups. All splits maintain strict pair-level integrity, meaning a vulnerability and its corresponding patch are always assigned to the same group.

To approximate familiarity with recurring vulnerability-fix patterns, patch-diff clusters are generated within each CWE class using MiniBatch K-Means on character-level TF-IDF features of the modified lines. Clustering the isolated diffs rather than full functions is essential: full-function representations would primarily group code by project style, variable naming, or surrounding framework context. In contrast, line-level diffs isolate the syntactic modification (e.g., operator adjustments, pointer validations) that resolves the flaw. Character-level n-grams are chosen because security patches often comprise short syntactic tokens, boundary operators, and numeric bounds rather than descriptive natural language. Furthermore, clustering is partitioned strictly per CWE with adaptive cluster budgets (e.g., 104 clusters for CWE-119 vs.\ 17 for CWE-415), preventing artificial over-fragmentation in sparse weakness categories while accounting for the heterogeneity of broader classes. The test set (845 pairs) is then structured as follows:

\begin{itemize}
    \item \textbf{Random ID (28.2\%)} A conventional held-out baseline drawn from the remaining in-distribution pool.
    \item \textbf{Seen-Pattern ID (18.9\%)} Unseen pairs whose CWE classes and specific patch-diff clusters were already observed during training.
    \item \textbf{Same-CWE New-Pattern OOD (21.1\%)} Isolates pattern-level generalization. The CWE class is known, but the specific patch-diff cluster is entirely withheld from training.
    \item \textbf{Project OOD (17.5\%)} Measures cross-project robustness by withholding complete repositories (e.g. krb5, postgres) from training \cite{steenhoek2023empirical}.
    \item \textbf{Temporal OOD (14.3\%)} Isolates later Common Vulnerabilities and Exposures (CVE) records (disclosed in 2024) to simulate future deployment and prevent temporal leakage \cite{jimenez2019importance}.
\end{itemize}

Evaluation groups are extracted through a strict priority cascade that withholds full projects first, then temporal CVEs from the remaining data, and finally unseen pattern clusters. This sequential partitioning prevents samples from satisfying multiple holdout conditions simultaneously, with automated checks verifying zero leakage across splits.

\subsection{Evaluation Metrics}
\label{subsec:evaluation_metrics}

Due to the balanced paired design, Accuracy serves as the primary classification metric, complemented by the F1 score. Because the negative class consists exclusively of repaired code, the Patch False-Positive Rate (Patch-FPR) is explicitly evaluated. A high Patch-FPR indicates that the model frequently flags repaired code as vulnerable, failing to recognize the security fix. Symmetrically, the False-Negative Rate (FNR) measures the proportion of genuine vulnerabilities that the model misses.

For OOD detection and distance-based risk, the Area Under the Receiver Operating Characteristic curve (AUROC) is used. Each sample receives a continuous distance score reflecting its geometric displacement from the training ID Core toward unfamiliar OOD regions (Fig.~\ref{fig:motivation_geometry}), formally defined in Section~\ref{subsec:ood_scoring}. AUROC evaluates the score's ability to rank unfamiliar samples or incorrect predictions higher than familiar ones, independent of an operational threshold \cite{fawcett2006introduction}.

\section{Methodology}
\label{sec:methodology}

The experimental design evaluates embedding calibration, OOD  scoring, and cross-model failure prediction. We examine two distinct model families, namely the CodeBERT encoder and the Qwen2.5-Coder language model. To isolate how task-specific training alters latent geometry compared to raw pretraining, we evaluate CodeBERT in both its original pretrained state and after fine-tuning with a joint cross-entropy and contrastive objective. The encoder produces fixed-dimensional embeddings for distance-based OOD scoring, while Qwen2.5-Coder acts as an independent generative classifier. This configuration allows us to test cross-model transfer, determining whether geometric distance scores derived from a lightweight encoder can reliably warn of prediction failures in a separate large language model.

\subsection{Model and Training Setup}
\label{subsec:model_setup}

The encoder architecture is based on CodeBERT (\texttt{microsoft/codebert-base})~\cite{codebert}. A central challenge in hard-negative vulnerability detection is that a vulnerable function and its patched counterpart share extensive syntactic overlap, yet a standard cross-entropy objective does not explicitly enforce geometric separation between their latent representations. To address this, the encoder is fine-tuned with a joint Cross-Entropy (CE) and Supervised Contrastive Learning (SCL) objective~\cite{gunel2020supervised, khosla2020supcon}. We adapted the contrastive loss formulation from framework implementations such as SCL-CVD and MCL-VD~\cite{sclcvd, mclvd}. Unlike approaches that fuse Abstract Syntax Trees (ASTs), control-flow graphs, or synthesized comments to optimize raw benchmark metrics, we restrict our encoder to unimodal source code. This eliminates parsing artifacts and confounders from external static analyzers, ensuring that observed geometric boundaries and neighbourhood structures directly reflect the latent properties of the contextualized language model itself. Following \cite{gunel2020supervised}, the optimization objective is defined as:
\[
  \mathcal{L} = \mathcal{L}_{\mathrm{CE}} + \lambda\,\mathcal{L}_{\mathrm{SCL}}
\]
where $\mathcal{L}_{\mathrm{CE}}$ is the standard cross-entropy classification loss. The contrastive term reshapes the embedding geometry directly by pulling same-label representations together and pushing differently labeled representations apart within each mini-batch:
\[
  \mathcal{L}_{\mathrm{SCL}} = \sum_{i \in I} \frac{-1}{|P(i)|} \sum_{p \in P(i)} \log \frac{\exp(\mathrm{sim}(z_i,z_p)/\tau)}{\sum_{a \in A(i)} \exp(\mathrm{sim}(z_i,z_a)/\tau)}
\]
where $\mathrm{sim}(\cdot,\cdot)$ denotes cosine similarity and $z_i = g(h_i) / \|g(h_i)\|_2$ represents the normalized projection of the sequence-level \texttt{[CLS]} token embedding $h_i$ through a projection head $g(\cdot)$. Here, $I \equiv \{1, \dots, N\}$ indexes the mini-batch, where $N=8$ represents the maximum batch size accommodated by GPU accelerator memory under full 512-token sequences. Ground-truth labels are denoted by $y_i \in \{0, 1\}$ (vulnerable vs.\ patched), $P(i) \equiv \{p \in A(i) \mid y_p = y_i\}$ is the set of positive indices sharing the anchor label, and $A(i) \equiv I \setminus \{i\}$. Following common practice in contrastive learning, the temperature is set to $\tau=0.07$~\cite{chen2020simclr, khosla2020supcon}, while, following the objective-balancing principles of Gunel et al.~\cite{gunel2020supervised} and SCL-CVD~\cite{sclcvd}, the scaling weight is set to $\lambda=0.05$. Training utilizes pair-complete mini-batches containing four exact vulnerable-patch pairs. This operationalizes contrastive hard-negative principles~\cite{robinson2021contrastive} by guaranteeing that the most informative counterpart, namely the repaired patch~\cite{xiao2020mvp}, is present in every optimization step despite the hardware-constrained batch size.

Both encoder variants are adapted with LoRA ($r=32$, $\alpha=64$, dropout $0.1$, applied to query, key, value, and dense projection modules)~\cite{hu2021lora}, building on recent empirical evidence showing that parameter-efficient adaptation preserves foundational representations while effectively capturing code task semantics~\cite{wang2025rag}. The final three transformer layers are additionally unfrozen, preserving general pretrained structure in lower layers while enabling task-specific abstraction in upper layers~\cite{lee2019freezing}.

As a cross-model reliability baseline, \texttt{Qwen2.5-Coder-7B-Instruct}~\cite{hui2024qwen25coder} is adapted as a sequence classifier via QLoRA (4-bit NF4, $r=32$, $\alpha=64$, dropout $0.05$)~\cite{dettmers2023qlora, jiang2025investigatingllmcvd, shestov2024finetuning}. The model input consists of a specific task instruction followed directly by the C or C++ code snippet: ``Analyze the following C/C++ code and determine if it contains a vulnerability.''

A binary classification head is applied to the final non-padding token of this combined sequence at the output layer. The LLM is fine-tuned using standard cross-entropy loss, where a pair-preserving sampler places vulnerable and patched counterparts within the same mini-batch to ensure balanced gradient updates across identical project contexts.

\subsection{Embedding Extraction \& Calibration}
\label{subsec:embedding_extraction}

Embeddings are extracted from the \texttt{[CLS]} token~\cite{bert} at every transformer layer (0--12) and as aggregated layer means. Layer choice is treated as an empirical factor because prior probing work shows a functional progression where  lower layers encode general syntax, while higher layers specialize toward the training objective~\cite{tenney2019bert}. A final layer tightly specialized for CE classification may lose the broader distributional structure required for OOD ranking.

Raw contextualized embeddings suffer from inherent anisotropy, clustering within a narrow cone of dominant directions~\cite{ethayarajh2019contextual}. In hard-negative vulnerability detection, where a function and its patch often differ by merely a few tokens (e.g., an added boundary check), this geometric distortion is critical: high-variance dimensions corresponding to surface syntax or project-specific identifier naming drown out the subtle semantic delta of the actual fix. To restore metric expressiveness, we apply a whitening transformation~\cite{su2021whitening} that centers and decorrelates the training distribution via the empirical covariance $\boldsymbol{\Sigma}=\mathbf{U}\boldsymbol{\Lambda}\mathbf{U}^\top$:
\[
  \mathbf{W} = (\boldsymbol{\Lambda} + \epsilon\mathbf{I})^{-1/2}\mathbf{U}^\top, \quad \mathbf{z}_{\mathrm{white}}(x) = \mathbf{W}\bigl(\mathbf{h}(x) - \boldsymbol{\mu}\bigr)
\]
where $\boldsymbol{\mu}$ is the training mean vector and $\epsilon=10^{-5}$ is a numerical regularizer. By equalizing eigenvalue variance across all feature dimensions, whitening prevents generic code structure from dominating cosine similarity, allowing localized repair semantics to govern latent distances.

\subsection{Distance-Based OOD Scoring}
\label{subsec:ood_scoring}

Vulnerability data is multi-modal, with different CWE classes, projects, and patch patterns occupying distinct local regions in the embedding space. Consequently, OOD scoring is operationalized using a non-parametric $k$-nearest-neighbour cosine distance to the training distribution~\cite{sun2022knnood}. Let $E(x)$ denote the calibrated embedding of sample $x$ under a given transformation (e.g., $E(x)=\mathbf{z}_{\mathrm{white}}(x)$), and let $N_k(x)$ denote its set of $k$ nearest training neighbours under cosine similarity. The non-parametric outlier score is computed as:
\[
  \mathrm{OOD}_k(x) = 1 - \frac{1}{k}\sum_{n_i \in N_k(x)} \cos\bigl(E(x),\, E(n_i)\bigr)
\]
The parameter $k$ is systematically swept across $k \in \{1, 3, 5, 10, 20, 30, 50, 100\}$. A diagonal approximation of the class-conditional Mahalanobis distance~\cite{lee2018mahalanobis} is included as a global parametric baseline. 

To evaluate whether distance scores expose a gradual generalization boundary, test samples are grouped into ten equal-frequency distance bins to compute accuracy and error rates. This is evaluated intrinsically (the encoder scored against its own predictions) and cross-model (using encoder-derived distances to rank LLM errors). Because neural networks can remain overconfident under distribution shift~\cite{guo2017calibration, ovadia2019trust}, the LLM's internal confidence (maximum softmax probability)~\cite{hendrycks2017baseline} is reported as a baseline. The analysis is repeated across shift types (in-distribution, unseen-pattern, cross-project, temporal), additionally measuring pair-exact correctness to assess whether both sides of a hard-negative pair are successfully identified.

\subsection{Vulnerable-Neighbourhood Proximity}
\label{subsec:proximity}

A specific hard-negative failure mode occurs when a patched function remains geometrically closer to vulnerable training examples than to safe counterparts, leading to misclassification. To investigate this, we analyze patched test samples using the fine-tuned CE+SCL encoder, contrasting patched false positives (predicted vulnerable) against patched true negatives (predicted patched). The contrastive objective makes this neighbourhood-based analysis directly interpretable by explicitly shaping class-wise structure~\cite{khosla2020supcon}.

To examine how test samples interact with the underlying training distribution, proximity to the vulnerable class is quantified using four complementary metrics in the calibrated embedding space. These include a centroid margin ratio, a nearest-neighbour margin, a class-conditional $k$-neighbourhood margin ratio ($k=10$), and the vulnerable-neighbour share among the top 10 neighbours illustrated in Fig.~\ref{fig:rq3_motivation_neighbourhood}. Grounded in neighbourhood-based prediction support and embedding similarity~\cite{papernot2018deepknn, shimmi2024vulsim}, class-conditional density estimation~\cite{lee2018mahalanobis}, and internal-state risk assessment~\cite{huang2025risk}, these metrics measure whether a patch is drawn toward the vulnerable or the safe training distribution. We assess their diagnostic value by comparing group means with 95\% confidence intervals, ranking false alarms via AUROC, and evaluating risk gradients across proximity deciles.

\section{Results}
\label{sec:results}

Table~\ref{tab:general_clf} summarizes baseline classification performance. Raw CodeBERT is excluded here as its head is unadapted to binary classification. It serves exclusively as a representation baseline.

\begin{table}[htbp]
\centering
\caption{Test classification performance. Patch-FPR denotes the fraction of patched samples incorrectly classified as vulnerable.}
\label{tab:general_clf}
\small
\setlength{\tabcolsep}{8pt}
\begin{tabular}{@{}lrrr@{}}
\toprule
\textbf{Model} & \textbf{Test Acc} & \textbf{Test F1} & \textbf{Patch-FPR} \\
\midrule
CodeBERT CE-only & 0.605 & 0.605 & 0.395 \\
CodeBERT CE+SCL  & 0.614 & 0.627 & 0.419 \\
Qwen2.5-Coder-7B & \textbf{0.627} & \textbf{0.635} & \textbf{0.394} \\
\bottomrule
\end{tabular}
\end{table}

All models perform above chance level, but none achieves high absolute performance. This aligns with prior findings that distinguishing vulnerabilities from their patches is considerably harder than standard classification \cite{xiao2020mvp}, where models often rely on superficial dataset cues rather than genuine defect semantics \cite{chakraborty2022deep}. Qwen2.5-Coder-7B achieves the strongest test accuracy and F1 score, but its margin over CodeBERT CE+SCL remains modest despite larger model capacity. At the same time, the patch false-positive rate remains consistently high across all models, demonstrating that classifiers capture vulnerable context while failing to verify whether the defect was repaired.

While Supervised Contrastive Learning (SCL) yields substantial gains in vision and NLP, its impact on raw binary classification here remains constrained. The bottleneck lies in the extreme syntactic overlap of hard-negative pairs: while SCL does not noticeably improve global OOD ranking over the raw encoder (RQ1), it successfully structures the latent space for local, class-conditional diagnostics of patched false positives (RQ3), even though translating this realignment into a superior linear decision boundary remains challenging.

\subsection{RQ1: Embedding Space Calibration}
\label{subsec:results_rq1}

RQ1 investigates representation choices for separating in-distribution and out-of-distribution samples, measured by mean OOD-AUROC across the three OOD conditions. 

The strongest overall configuration across the grid uses raw pretrained CodeBERT at layer 3 with whitening and $k=1$, achieving a mean OOD-AUROC of 0.739. Notably, this best geometry stems from the raw encoder rather than fine-tuned variants (which peak at 0.735 for CE+SCL), indicating that fine-tuning optimizes decision boundaries at the expense of broader OOD geometry. Post-whitening L2 normalization yielded identical results, hence only whitening is reported.

\paragraph{Effect of Embedding Transformation}
Raw embeddings with cosine nearest-neighbour scoring reach a best mean OOD-AUROC of 0.710, whereas whitened variants reach 0.739 (Table~\ref{tab:rq1_transform}), confirming that decorrelating dominant variance directions restores metric utility. The failure of the diagonal Mahalanobis baseline (best mean 0.547) underscores that code repositories and CWE mechanisms form a fragmented multi-modal manifold rather than a single Gaussian cluster. Consequently, distance scoring strictly favors local nearest neighbours ($k=1$ achieves 0.739, as shown later in Table~\ref{tab:rq1_k}), as snippets are familiar if they match even a single recurring implementation pattern in training.

\begin{table}[htbp]
\centering
\caption{Effect of embedding transformation and scoring method on OOD separability (aggregated over all layers and model variants).}
\label{tab:rq1_transform}
\small
\setlength{\tabcolsep}{2.5pt}
\begin{tabular}{@{}llrr@{}}
\toprule
\textbf{Transformation} & \textbf{Method} & \textbf{Best Mean} & \textbf{Avg.\ Mean} \\
\midrule
Whitened    & cosine $k$NN & \textbf{0.739} & \textbf{0.613} \\
Raw         & cosine $k$NN & 0.710 & 0.576 \\
Raw         & Mahalanobis  & 0.547 & 0.513 \\
\bottomrule
\end{tabular}
\end{table}

\paragraph{Effect of Encoder Layer}
As shown in Table~\ref{tab:rq1_layers}, non-contextual token embeddings (layer 0) carry zero distributional signal (AUROC 0.500). OOD separability emerges immediately upon contextualization (layer 1, 0.711) and peaks in early-to-intermediate layers (layers 3--5, peaking at 0.739 in layer 3). In final layers (10--12), task-specific fine-tuning pressure forcibly collapses structural contexts onto the classification boundary, discarding variance needed for OOD estimation. Layer 3 retains rich contextual syntax without objective-driven collapse, providing the optimal substrate for distance scoring.

\begin{table}[htbp]
\centering
\caption{Layer-wise OOD separability. Intermediate layers retain the strongest distributional signal.}
\label{tab:rq1_layers}
\small
\setlength{\tabcolsep}{4pt}
\begin{tabular}{@{}lrrrr@{}}
\toprule
\textbf{Layer} & \textbf{Best Mean} & \textbf{Pattern} & \textbf{Project} & \textbf{Time} \\
\midrule
0  & 0.500 & 0.500 & 0.500 & 0.500 \\
1  & 0.711 & 0.668 & 0.750 & 0.758 \\
2  & 0.720 & 0.685 & 0.746 & 0.779 \\
3  & \textbf{0.739} & 0.697 & \textbf{0.776} & 0.793 \\
4  & 0.734 & 0.699 & 0.774 & \textbf{0.794} \\
5  & 0.735 & 0.706 & 0.774 & 0.790 \\
6  & 0.721 & 0.708 & 0.748 & 0.778 \\
7  & 0.718 & 0.711 & 0.725 & 0.782 \\
8  & 0.700 & 0.711 & 0.718 & 0.743 \\
9  & 0.708 & 0.712 & 0.731 & 0.743 \\
10 & 0.704 & \textbf{0.718} & 0.726 & 0.732 \\
11 & 0.717 & 0.709 & 0.727 & 0.750 \\
12 & 0.708 & 0.706 & 0.748 & 0.729 \\
\bottomrule
\end{tabular}
\end{table}

\paragraph{Effect of Neighbourhood Size ($k$)}
The OOD signal is strictly local (Table~\ref{tab:rq1_k}). A single nearest neighbour ($k=1$) yields the strongest mean AUROC (0.739), which degrades monotonically as $k$ expands due to cluster mixing. Project OOD is the sole exception, peaking at $k=3$ (0.776), consistent with project conventions forming small cohesive neighbourhoods.

\begin{table}[htbp]
\centering
\caption{Effect of neighbourhood size ($k$) on cosine-$k$NN OOD separability. Group columns report marginal maxima.}
\label{tab:rq1_k}
\small
\setlength{\tabcolsep}{4pt}
\begin{tabular}{@{}lrrrr@{}}
\toprule
\textbf{$k$} & \textbf{Best Mean} & \textbf{Pattern} & \textbf{Project} & \textbf{Time} \\
\midrule
1   & \textbf{0.739} & \textbf{0.720} & 0.745 & \textbf{0.794} \\
3   & 0.721 & 0.693 & \textbf{0.776} & 0.740 \\
5   & 0.703 & 0.673 & 0.774 & 0.735 \\
10  & 0.676 & 0.659 & 0.750 & 0.709 \\
20  & 0.650 & 0.623 & 0.721 & 0.676 \\
30  & 0.631 & 0.609 & 0.713 & 0.651 \\
50  & 0.611 & 0.597 & 0.706 & 0.622 \\
100 & 0.598 & 0.593 & 0.690 & 0.589 \\
\bottomrule
\end{tabular}
\end{table}

\begin{tcolorbox}[title=\textbf{Answer to RQ1}, colback=gray!5, colframe=gray!60]
Raw pretrained CodeBERT (Layer 3) with whitening calibration and local cosine scoring ($k=1$) achieves the best OOD separability (mean AUROC 0.739). While fine-tuning optimizes classification, representation geometry for detecting shift is best preserved in intermediate layers prior to task-specific collapse.
\end{tcolorbox}

\subsection{RQ2: Generalization Boundaries}
\label{subsec:results_rq2}

RQ2 investigates whether embedding-based distance exposes a reliability gradient and whether this gradient behaves uniformly across distribution shifts. All following analyses use the layer-3 whitened configuration ($k=1$) identified in RQ1. For interpretability, the test set is divided into ten equal-frequency distance bins (169 samples each) wherever a bin-level analysis is reported.

\begin{figure}[t]
\centering
\includegraphics[width=0.82\linewidth]{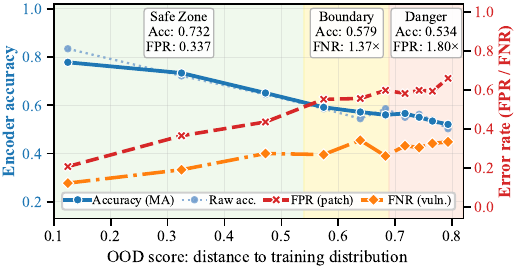}
\caption{CodeBERT CE+SCL encoder performance over its own embedding distance (ten equal-frequency bins). Accuracy decreases and patch FPR generally rises with increasing distance from the training distribution.}
\Description{Line graph showing CodeBERT accuracy dropping from 83.4\% to 47.9\% and patch false positive rate rising from 16.7\% to 63.1\% across ten equal-frequency embedding distance bins.}
\label{fig:rq2_encoder_boundary}
\end{figure}

\paragraph{Encoder-Intrinsic and Cross-Model Boundaries}
Figure~\ref{fig:rq2_encoder_boundary} shows CodeBERT CE+SCL performance over its own embedding distance. A clear degradation pattern emerges. Accuracy falls from 0.834 in the lowest-distance bin to 0.479 in the highest, which is a drop of more than 35 percentage points, while the patch false-positive rate rises from 0.167 to 0.631. Converting prediction correctness into a binary error indicator, the resulting error-ranking AUROC is 0.593 (Spearman $r=0.157$). This confirms a modest but meaningful reliability signal consistent with the large accuracy drop (full ten-bin results available in the online replication package).

Operationally, this geometric gradient enables prediction and risk-aware triage in software security pipelines. By grouping distance bins into empirical zones—Safe (Bins 1--3), Boundary (Bins 4--6), and Danger (Bins 7--10)—a system can selectively abstain from fully automated decisions on distant code. Abstaining on samples in the highest-distance Danger Zone yields 71.6\% accuracy on the retained low-distance code (Safe Zone) compared to 53.4\% in the rejected region (Danger Zone), while reducing the false alarm rate on safe code down to 33.7\%.

\begin{figure}[t]
\centering
\includegraphics[width=0.82\linewidth]{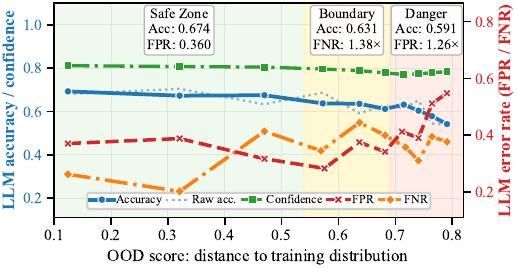}
\caption{Qwen2.5-Coder performance over raw CodeBERT embedding distance. Accuracy decreases with distance, while mean confidence remains comparatively stable.}
\Description{Line graph showing Qwen2.5-Coder LLM accuracy decreasing from 68.0\% to 53.9\% across distance bins, while LLM confidence remains comparatively stable around 80\%.}
\label{fig:rq2_llm_boundary}
\end{figure}

To evaluate whether this geometric signal transfers across model families, Figure~\ref{fig:rq2_llm_boundary} uses raw CodeBERT distance to rank errors made by the independently fine-tuned Qwen2.5-Coder LLM. LLM accuracy decreases from 0.680 to 0.539 across distance bins, while mean confidence declines only slightly from $\approx$0.81 to $\approx$0.78. This reproduces the known overconfidence of neural networks under distribution shift~\cite{guo2017calibration, ovadia2019trust}. While the cross-model error-ranking AUROC (0.556) is weaker than the encoder-intrinsic setting, it remains consistently above random ranking. CodeBERT distance and LLM confidence thus capture complementary reliability aspects: confidence reflects internal certainty, while distance provides an external measure of distributional proximity.

\begin{figure}[t]
\centering
\includegraphics[width=0.7\linewidth]{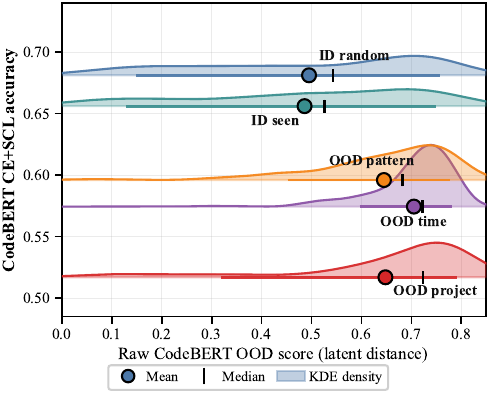}
\caption{Shift-specific boundary behaviour and decoupling. Group centroids (circles) and medians (vertical bars) relate raw CodeBERT OOD distance to CE+SCL accuracy (Y-axis), while horizontal ridge curves display the OOD score density for each group. Despite temporal shift drifting furthest, cross-project shift causes the sharpest accuracy degradation.}
\Description{2D ridge and scatter plot relating embedding distance on the horizontal axis to accuracy on the vertical axis across five evaluation groups, showing cross-project shift causing the lowest accuracy despite temporal shift having higher distance.}
\label{fig:rq3_density}
\end{figure}

\paragraph{Shift-Specific Boundary Behaviour}
Figure~\ref{fig:rq3_density} relates distributional displacement to classification degradation across the evaluation groups. The two in-distribution baselines remain clustered in lower distance regions ($\approx$0.49), while pattern and project shifts displace samples outward ($\approx$0.65). Temporal samples exhibit the highest mean distance (0.705) alongside a tightly bounded variance (standard deviation of 0.080).

Across shifts, geometric distance and classification degradation decouple. Temporal samples are geometrically farthest from the training distribution, yet CE+SCL encoder accuracy on temporal samples (0.574) exceeds that on the project group (0.517). The project group represents the most severe failure mode overall, suffering the lowest accuracy alongside the highest false-positive rate (0.507), meaning more than half of all patched cross-project samples are misclassified as vulnerable. Temporal shift instead produces the highest false-negative rate (0.455). This demonstrates that high distance is a necessary but not sufficient condition for severe degradation. The structural nature of the shift determines how displacement translates into specific failure modes.

This non-linear divergence is grounded in representation geometry. While temporal shift induces a uniform distributional drift away from the training corpus across all classes, cross-project shift causes severe representation entanglement. As neighbourhood diagnostics reveal (detailed in the supplementary package), cross-project patched samples are the only evaluation group exhibiting a positive group-level centroid vulnerability margin (+0.016, detailed in the supplementary repository). Unseen repository conventions, custom macros, and unfamiliar API patterns do not merely push samples outward—they actively displace safe patches into the geometric sphere of vulnerable training clusters, triggering systematic false alarms (see supplementary material).

\begin{figure}[htbp]
\centering
\includegraphics[width=0.82\linewidth]{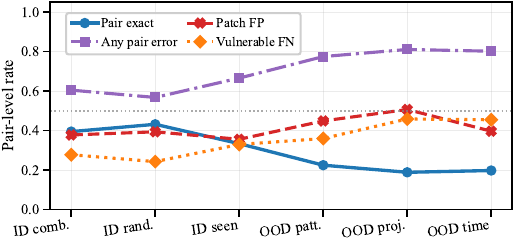}
\caption{Hard-negative pair behaviour (CE+SCL encoder). Pair-exact correctness decreases sharply under OOD shifts, exposing shift-specific failure modes on either the patch (FPR) or the vulnerability (FNR).}
\Description{Line chart showing pair-exact correctness dropping below the 25\% random threshold under OOD shifts, while patch false positive rate peaks on project shift and vulnerable false negative rate peaks on temporal shift.}
\label{fig:rq3_pair_metrics}
\end{figure}

\paragraph{Hard-Negative Pair Behaviour}
Evaluating the same shifts at the level of strictly vulnerable--patched pairs exposes a more severe failure picture (Figure~\ref{fig:rq3_pair_metrics}). Even in the combined in-distribution setting, only 39.5\% of pairs are classified exactly correctly (where both the vulnerability and its patch are identified correctly). For an independent binary classifier, predicting both sides of a pair correctly yields an expected random baseline of 25\% ($0.5 \times 0.5$). Under OOD shifts, pair-exact correctness consistently drops below this threshold: 22.5\% for unseen patterns, 19.8\% for temporal shifts, and 18.9\% across projects. This demonstrates that models under shift do not merely degrade gradually, but systematically misclassify the localized repair delta. These shift-specific paired error profiles are invisible at the sample level, confirming that pair-aware evaluation is essential for hard-negative reliability.

\paragraph{CWE-Level Failure Modes}
Beyond shift type, generalization boundaries depend heavily on the specific vulnerability class. Table~\ref{tab:rq2_cwe} reports performance per CWE for the CE+SCL encoder (full class breakdown in supplementary repository). CWE-787 (Out-of-bounds Write) is the most robust class with the highest accuracy (0.722). In contrast, CWE-416 (Use After Free) is the weakest overall (accuracy 0.542) despite substantial training support (1,636 samples) and a moderate mean OOD distance (0.566). This indicates that poor performance is intrinsic to the semantic complexity of the class rather than a pure distance effect. The dominant error type also varies across classes. CWE-415 (Double Free) suffers from a severe false-negative rate (0.528), plausibly compounded by limited training support. Conversely, CWE-476 (NULL Pointer Dereference) reaches the highest mean OOD distance (0.659) alongside a high false-positive rate (0.483). Vulnerability semantics therefore modulate the generalization boundary independently of geometric distance and shift type.

\begin{table}[htbp]
\centering
\caption{CWE-level failure modes for the CE+SCL encoder. Performance varies heavily by vulnerability semantics, independent of embedding distance. Snippets may carry multiple CWE labels, thus counts do not sum to total test size.}
\label{tab:rq2_cwe}
\small
\setlength{\tabcolsep}{4.5pt}
\begin{tabular}{@{}lrrrrrr@{}}
\toprule
\textbf{CWE} & \textbf{Test $n$} & \textbf{OOD} & \textbf{Acc} & \textbf{FPR} & \textbf{FNR} & \textbf{Error} \\
\midrule
CWE-787 & 234 & \textbf{0.546} & \textbf{0.722} & \textbf{0.316} & \textbf{0.239} & \textbf{0.278} \\
CWE-120 & 82  & 0.595 & 0.671 & 0.390 & 0.268 & 0.329 \\
CWE-125 & 290 & 0.557 & 0.669 & 0.379 & 0.283 & 0.331 \\
CWE-119 & 500 & 0.566 & 0.596 & 0.428 & 0.380 & 0.404 \\
CWE-476 & 290 & \textbf{0.659} & 0.600 & \textbf{0.483} & 0.317 & 0.400 \\
CWE-415 & 106 & 0.655 & 0.557 & 0.358 & \textbf{0.528} & 0.443 \\
CWE-416 & 240 & 0.566 & \textbf{0.542} & 0.475 & 0.442 & \textbf{0.458} \\
\bottomrule
\end{tabular}
\end{table}

\begin{tcolorbox}[title=\textbf{Answer to RQ2}, colback=gray!5, colframe=gray!60]
Embedding distance reveals a gradual, shift-specific generalization boundary. Within the encoder, accuracy drops from 0.834 to 0.479 as distance increases. This signal transfers weakly to an independent LLM, exposing its overconfidence on distant samples. However, distance alone does not determine failure severity. Temporal samples are geometrically farthest but cause less degradation than cross-project shifts. Pair-level and CWE-level analyses confirm that vulnerability semantics modulate the boundary independently of embedding distance.
\end{tcolorbox}

\subsection{RQ3: Vulnerable-Neighbourhood Proximity and Patched False Positives}
\label{subsec:results_rq3}

RQ3 investigates whether patched false positives are geometrically closer to vulnerable regions of the training embedding space than correctly classified patched samples. The analysis uses the CE+SCL layer-3 whitened embeddings identified in RQ1, since these are the representations of the classifier that actually produces the observed patched false positives. The evaluated subset consists exclusively of patched test samples, comparing patched false positives (FPs) against patched true negatives (TNs).

Across all proximity metrics evaluated in Table~\ref{tab:rq3_proximity}, patched false positives reside systematically closer to vulnerable training neighbourhoods than true negatives. For instance, safe patches misclassified as vulnerable contain 51.1\% vulnerable training samples in their top-10 neighbourhood versus 46.0\% for true negatives, peaking at an AUROC of 0.636 (Spearman $r=0.232$) for the $k=10$ distance ratio. All 95\% confidence intervals strictly exclude zero, confirming stable representation entanglement across the test set.

\begin{table}[htbp]
\centering
\caption{RQ3 proximity analysis for patched FPs vs.\ TNs ($\Delta$ = FP minus TN; all $p < 10^{-5}$, two-sided Welch's $t$-test). For ratios, lower values indicate stronger vulnerable proximity.}
\label{tab:rq3_proximity}
\small
\setlength{\tabcolsep}{5pt}
\begin{tabular}{@{}lrrrr@{}}
\toprule
\textbf{Metric} & \textbf{FP} & \textbf{TN} & \textbf{$\Delta$ [95\% CI]} & \textbf{AUROC} \\
\midrule
Centroid margin       &  0.009 & -0.030 &  0.039 [0.027, 0.050]   & 0.609 \\
NN margin             & -0.012 & -0.040 &  0.029 [0.018, 0.040]   & 0.591 \\
Vuln.\ share ($k$=10) &  0.511 &  0.460 &  0.051 [0.035, 0.068]   & 0.597 \\
Class margin ($k$=10) & -0.001 & -0.017 &  0.016 [0.012, 0.021]   & 0.635 \\
Dist.\ ratio ($k$=10) &  1.002 &  1.032 & -0.030 [-0.041, -0.021] & \textbf{0.636} \\
\bottomrule
\end{tabular}
\end{table}

\begin{figure}[t]
\centering
\includegraphics[width=0.82\linewidth]{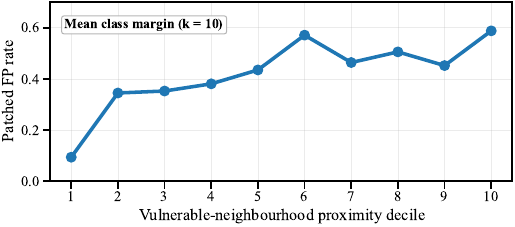}
\caption{Patched FP rate across proximity-risk deciles (class-specific $k=10$ vulnerability margin). Higher deciles indicate greater proximity to vulnerable training neighbourhoods.}
\Description{Line chart of patched false positive rate across ten proximity deciles, surging from 9.4\% in the lowest decile to 58.8\% in the highest decile.}
\label{fig:rq4_fp_decile}
\end{figure}

Figure~\ref{fig:rq4_fp_decile} confirms this risk gradient: the patched FP rate surges from 9.4\% in the lowest proximity decile to 58.8\% in the highest. While intermediate deciles show slight non-monotonicity, vulnerable-neighbourhood proximity effectively isolates severe false-alarm zones.

A group-level breakdown (see supplementary repository) connects this proximity effect to the shift-specific findings from RQ2. The cross-project group is the only group with a positive centroid vulnerability margin (+0.016), indicating that patched cross-project samples are geometrically closer to the vulnerable centroid than to the patched centroid. This provides a representation-space explanation for the over-warning failure mode identified in RQ2. Cross-project shift moves patched code into embedding regions that resemble the vulnerable class, irrespective of whether the repair has actually been applied. 

\begin{tcolorbox}[title=\textbf{Answer to RQ3}, colback=gray!5, colframe=gray!60,]
Patched false positives are associated with geometric proximity to vulnerable training neighbourhoods. The $k=10$ distance ratio ranks these errors with an AUROC of 0.636, while the false positive rate increases from 9.4\% in the lowest to 58.8\% in the highest proximity-risk decile. Local representation geometry thus provides an runtime signal to triage and flag high-risk predictions before deployment.
\end{tcolorbox}

\section{Discussion}
\label{sec:discussion}

The results support the argument that calibrated embedding distances make generalization boundaries empirically observable, while revealing qualifications about when and how this signal is useful.

\subsection{Interpretation of the Main Findings}
\label{subsec:discussion_main_findings}

\paragraph{Classification Performance and Representation Geometry}
The decoupling between classification accuracy and OOD separability observed in RQ1 suggests that fine-tuning reshapes the embedding space toward the classification boundary at the expense of the broader distributional structure needed for reliable OOD ranking~\cite{kumar2022finetuningdistort}. This decoupling carries direct architectural implications for secure software engineering pipelines. Standard practice assumes that safety checks and confidence scores should originate from the final layer of the active classifier. Our findings demonstrate that task fine-tuning forces an objective-driven feature collapse in upper layers (10--12) to linearly separate labels, destroying the broader contextual variance required for OOD scoring. Synthesizing these layer dynamics with our cross-model transfer results in RQ2, production security tools should adopt a dual-representation architecture. A high-capacity classifier such as a fine-tuned LLM performs primary vulnerability triage, while an auxiliary, frozen intermediate encoder representation (CodeBERT Layer 3 with whitening) operates as an independent guardrail to flag out-of-distribution inputs. This dual architecture adds negligible latency to the screening pipeline. With 125M parameters, the frozen CodeBERT encoder is over 50 times smaller than a 7B LLM and computes distance scores in under 5\,ms per function, allowing security tools to screen code before invoking resource-heavy generative inference or manual triage.

\paragraph{Distance-Based Boundaries and Model Dependence}
That LLM confidence remains comparatively high even as accuracy declines with distance is consistent with broader evidence that neural-network confidence is poorly calibrated and degrades further under distribution shift~\cite{guo2017calibration, ovadia2019trust}. This non-redundancy has a direct practical consequence: a system relying on model confidence alone would fail to flag exactly the high-distance region where CodeBERT-derived distance proves informative.

\paragraph{Shift-Specific and Geometric Failure Modes}
The RQ3 proximity analysis offers a representation-level account of why cross-project shift is harder than its distance alone would predict~\cite{steenhoek2023empirical}: it is the only evaluation group whose patched samples are geometrically closer to the vulnerable class centroid than to the patched class. While prior work established that representation geometry and neighbour embeddings capture vulnerability similarity~\cite{shimmi2024vulsim}, our findings ground shift-induced failures directly in this geometry, confirming that these errors reflect measurable shortcut learning~\cite{chakraborty2022deep}.

\paragraph{Vulnerability-Class-Dependent Failure Modes} The CWE-level results show that vulnerability semantics shape the generalization boundary beyond distance and shift type. CWE-416 remains the weakest class despite substantial training support, suggesting that lifetime and aliasing relations are difficult to recover from short truncated snippets. CWE-415 shows a similar false-negative pattern, whereas CWE-787 and CWE-120 appear more robust because their signals are more locally observable. Overall, lifetime- and control-flow-dependent weaknesses likely require representations beyond token-level context.

\subsection{Threats to Validity}
\label{subsec:discussion_threats}

\paragraph{Construct Validity}
Using vulnerable--patched pairs isolates the security delta but omits arbitrary benign code common in deployment. Furthermore, while hunk extraction ensures the root cause fits within the model's context window, real-world tools must inspect entire functions where the vulnerability location is unknown \textit{a priori}. Finally, patch-diff clusters provide an automated operational proxy for fix categories rather than manually validated semantic taxonomies.

\paragraph{Internal Validity}
Pair integrity, leakage prevention, and duplicate removal strengthen internal validity. Source datasets may contain label noise or non-security refactoring changes within patches, influencing both classification and distance. The character-level input limit and hunk extraction mitigate arbitrary tokenizer truncation but may remove broader function context relevant to specific vulnerabilities. Additionally, RQ2 and RQ3 condition their analyses on the strongest configuration identified in RQ1 (full parameter grid in supplementary material).

\paragraph{External Validity}
The dataset is restricted to seven memory and resource-safety CWE classes in C and C++. This supports controlled hard-negative analysis but limits generalization to web vulnerabilities, cryptographic misuse, or other programming languages. The evaluation relies on specific held-out projects and a 2024 temporal cutoff. The evaluated models are limited to CodeBERT and Qwen2.5-Coder-7B. Results may differ for encoders with explicit data-flow inputs or significantly larger LLM variants.

\paragraph{Conclusion Validity}
Model comparisons rely on fixed training runs rather than extensive repeated-run studies. The conclusions emphasize consistent qualitative patterns across evaluation groups rather than minor absolute metric differences. The equal-frequency distance bins used for boundary analysis are empirical and dataset-dependent. Statistical association does not establish strict causality. Proximity to vulnerable neighbourhoods is a geometric contributor to patched false positives, but it interacts with local classifier thresholds, input truncation, and project context.

\section{Related Work}
\label{sec:related_work}
The literature relevant to this study spans four areas.
\paragraph{Distribution Shift in Vulnerability Detection}
Chakraborty et al.~\cite{chakraborty2022deep} showed that deep detectors rely on shortcut features that fail under distribution shift. Jimenez et al.~\cite{jimenez2019importance} identified temporal leakage as a distinct cause, with recent studies confirming that LLMs continue to struggle under software evolution and dataset shifts over time \cite{bhutamapuram2025llms}. Steenhoek et al.~\cite{steenhoek2023empirical} isolated project identity as a further factor in model degradation, underscoring ongoing efforts to transition vulnerability benchmarks from isolated functions to realistic repository-level contexts~\cite{lbath2026toward}. Li et al.~\cite{li2025outofdistribution} confirmed in a controlled benchmark that in-distribution performance does not predict OOD generalization. These studies largely treat OOD code as a single undifferentiated category. The present work extends their methodology by systematically isolating pattern, project, and temporal shifts across all evaluations. Our RQ2 findings reveal that geometric distance and classification failure do not correlate linearly across shift categories, showing that project-level shift degrades model reliability far out of proportion to its distributional displacement.

\paragraph{Hard-Negative and Patch-Based Detection}
Xiao et al.~\cite{xiao2020mvp} combined vulnerability and patch signatures in MVP to avoid flagging already-repaired code. While recent research focuses on prioritizing alerts and mitigating false alarms in automated tools~\cite{amusuo2026lessons}, deep learning detectors themselves still suffer from high false-alarm rates when inspecting repaired functions. Fu and Tantithamthavorn~\cite{fu2022linevul} showed with LineVul that vulnerability evidence is often confined to a few lines, explaining why patched functions form strict hard-negatives. In parallel, SCL-CVD and MCL-VD~\cite{sclcvd, mclvd} pursue class separation via contrastive training, while Shimmi et al.~\cite{shimmi2024vulsim} and Papernot and McDaniel~\cite{papernot2018deepknn} demonstrated that local neighbourhood structure carries diagnostic value. While prior research uses hard-negatives and neighbourhoods mainly as training or retrieval mechanisms, this paper applies them diagnostically. Our pair-exact evaluation confirms the concerns behind MVP directly, and our proximity analysis gives the shortcut behaviour anticipated by prior work a concrete geometric signature.

\paragraph{Contextual Embedding Geometry}
Ethayarajh~\cite{ethayarajh2019contextual} established that contextual embeddings are inherently anisotropic. Su et al.~\cite{su2021whitening} and Huang et al.~\cite{huang2021whiteningbert} showed that whitening restores distance as a meaningful similarity measure for sentence embeddings, and Tenney et al.~\cite{tenney2019bert} demonstrated via edge probing that transformer layers specialize progressively with depth. The findings in RQ1 confirm these concepts jointly for code representations. Whitening effectively corrects for anisotropy, while the layer-wise analysis extends depth progression to an OOD setting. Our results show that the last-layer specialization described in prior work comes at a direct cost to the broader distributional structure required for reliable distance-based analysis.

\paragraph{OOD Detection and Model Calibration}
Hendrycks and Gimpel~\cite{hendrycks2017baseline} established maximum softmax probability as an OOD baseline. Guo et al.~\cite{guo2017calibration} showed modern networks are poorly calibrated, and Ovadia et al.~\cite{ovadia2019trust} demonstrated this worsens under dataset shift. For geometric scoring, Lee et al.~\cite{lee2018mahalanobis} proposed a parametric class-conditional alternative, while Sun et al.~\cite{sun2022knnood} proposed a non-parametric $k$NN distance. Our calibration results confirm Sun et al.'s findings for code, demonstrating that the most useful OOD signal is local rather than global. Furthermore, the cross-model evaluation confirms the overconfidence problem for code-specific LLMs, showing that an externally calibrated encoder distance exposes reliability information that the LLM's own confidence fails to capture, reinforcing findings on the unreliability of raw model confidence in software diagnostics~\cite{zhang2024lmpace}.

\section{Conclusion}
\label{sec:conclusion}

This study demonstrates that calibrated embedding distances make the generalization boundaries of vulnerability detection models empirically observable under hard-negative conditions. Scaling model capacity alone does not resolve hard-negative confusion, as generative LLMs remain vulnerable to the same localized syntax traps as smaller encoders. We show that whitened intermediate CodeBERT layers provide the strongest separation of familiar and unfamiliar code. Distance-based boundaries are gradual and shift-specific: cross-project shifts cause the sharpest accuracy drops, while temporal shifts push samples furthest from the training distribution, heavily inflating false negatives. Patched false positives systematically correlate with geometric proximity to vulnerable training neighbourhoods.

\textbf{Practical implications} include the ability to retrofit existing security scanners directly with out-of-distribution scores derived from intermediate layers via post-hoc whitening. Because this transformation requires no gradient updates or separate shadow models, practitioners can introduce dynamic risk stratification into CI/CD pipelines with negligible compute overhead. This enables selective prediction: automated tools can confidently flag vulnerabilities in familiar code regions while routing geometrically distant or entangled functions to security analysts. Because error profiles are shift-dependent, distance scores indicate which operational trade-off to budget for: auditing unfamiliar repositories requires reviewer capacity to handle inflated false alarms, whereas inspecting temporally shifted code necessitates recall-focused monitoring against silently missed defects. 

\textbf{Future work} should extend this geometric framework beyond textual representations to graph- and data-flow-based models, which may better capture lifetime-dependent vulnerabilities such as Use-After-Free flaws. In addition, future research should explore explicit fine-tuning objectives tailored specifically to OOD density estimation rather than relying solely on cross-entropy classification, and validate boundary stability across multi-year longitudinal code evolution.

\section*{Data Availability}
All datasets, evaluation splits, model checkpoints, and reproduction notebooks are accessible at \url{https://figshare.com/s/2a9e27691dbf03990ece}.

\bibliographystyle{ACM-Reference-Format}
\bibliography{literature}

@inproceedings{allamanis2019adverse,
  author      = {Miltiadis Allamanis},
  title       = {The Adverse Effects of Code Duplication in Machine Learning Models of Code},
  booktitle   = {Proceedings of the 2019 ACM SIGPLAN International Symposium on New Ideas, New Paradigms, and Reflections on Programming and Software (Onward! 2019)},
  pages       = {143--153},
  publisher   = {Association for Computing Machinery},
  year        = {2019},
  doi         = {10.1145/3359591.3359735}
}

@inproceedings{amusuo2026lessons,
  author      = {Paschal C. Amusuo and
                  Dongge Liu and
                  Ricardo Andres Calvo Mendez and
                  Jonathan Metzman and
                  Oliver Chang and
                  James C. Davis},
  title       = {Lessons from Mitigating False Positives in Google's {OSS-Fuzz-Gen}},
  booktitle   = {Proceedings of the 34th ACM International Conference on the Foundations of Software Engineering (FSE Companion '26)},
  pages       = {316--321},
  publisher   = {Association for Computing Machinery},
  year        = {2026},
  doi         = {10.1145/3803437.3805205}
}

@inproceedings{bert,
  author      = {Jacob Devlin and
                  Ming-Wei Chang and
                  Kenton Lee and
                  Kristina Toutanova},
  title       = {{BERT}: Pre-training of Deep Bidirectional Transformers for Language Understanding},
  booktitle   = {Proceedings of the 2019 Conference of the North American Chapter of the Association for Computational Linguistics: Human Language Technologies (NAACL-HLT 2019)},
  pages       = {4171--4186},
  publisher   = {Association for Computational Linguistics},
  year        = {2019},
  doi         = {10.18653/v1/N19-1423}
}

@inproceedings{bhutamapuram2025llms,
  author      = {Umamaheswara Sharma Bhutamapuram and
                  Farhan Chonari and
                  Gokul K. Anilkumar and
                  Sai Kiran Konchada},
  title       = {{LLMs} for Defect Prediction in Evolving Datasets: Emerging Results and Future Directions},
  booktitle   = {Proceedings of the 33rd ACM International Conference on the Foundations of Software Engineering (FSE Companion '25)},
  pages       = {520--524},
  publisher   = {Association for Computing Machinery},
  year        = {2025},
  doi         = {10.1145/3696630.3728491}
}

@article{chakraborty2022deep,
  author      = {Saikat Chakraborty and
                  Rahul Krishna and
                  Yangruibo Ding and
                  Baishakhi Ray},
  title       = {Deep Learning Based Vulnerability Detection: Are We There Yet?},
  journal     = {IEEE Transactions on Software Engineering},
  volume      = {48},
  number      = {9},
  pages       = {3280--3296},
  publisher   = {IEEE Computer Society},
  year        = {2022},
  doi         = {10.1109/TSE.2021.3087402}
}

@inproceedings{chen2020simclr,
  author      = {Ting Chen and
                  Simon Kornblith and
                  Mohammad Norouzi and
                  Geoffrey Hinton},
  title       = {A Simple Framework for Contrastive Learning of Visual Representations},
  booktitle   = {Proceedings of the 37th International Conference on Machine Learning (ICML 2020)},
  series      = {Proceedings of Machine Learning Research},
  volume      = {119},
  pages       = {1597--1607},
  publisher   = {PMLR},
  year        = {2020},
  url         = {http://proceedings.mlr.press/v119/chen20j.html}
}

@inproceedings{codebert,
  author      = {Zhangyin Feng and
                  Daya Guo and
                  Duyu Tang and
                  Nan Duan and
                  Xiaocheng Feng and
                  Ming Gong and
                  Linjun Shou and
                  Bing Qin and
                  Ting Liu and
                  Daxin Jiang and
                  Ming Zhou},
  title       = {{CodeBERT}: A Pre-Trained Model for Programming and Natural Languages},
  booktitle   = {Findings of the Association for Computational Linguistics: EMNLP 2020},
  pages       = {1536--1547},
  publisher   = {Association for Computational Linguistics},
  year        = {2020},
  doi         = {10.18653/v1/2020.findings-emnlp.139}
}

@inproceedings{croft2023dataquality,
  author      = {Roland Croft and
                  M. Ali Babar and
                  M. Mehdi Kholoosi},
  title       = {Data Quality for Software Vulnerability Datasets},
  booktitle   = {Proceedings of the 2023 IEEE/ACM 45th International Conference on Software Engineering (ICSE '23)},
  pages       = {121--133},
  publisher   = {IEEE Computer Society},
  year        = {2023},
  doi         = {10.1109/ICSE48619.2023.00022}
}

@inproceedings{dettmers2023qlora,
  author      = {Tim Dettmers and
                  Artidoro Pagnoni and
                  Ari Holtzman and
                  Luke Zettlemoyer},
  title       = {{QLoRA}: Efficient Finetuning of Quantized {LLMs}},
  booktitle   = {Advances in Neural Information Processing Systems (NeurIPS 2023)},
  volume      = {36},
  pages       = {10088--10115},
  publisher   = {Curran Associates, Inc.},
  year        = {2023},
  doi         = {10.52202/075280-0441}
}

@inproceedings{ding2024vulnerability,
  author      = {Yangruibo Ding and
                  Yanjun Fu and
                  Omniyyah Ibrahim and
                  Chawin Sitawarin and
                  Xinyun Chen and
                  Basel Alomair and
                  David Wagner and
                  Baishakhi Ray and
                  Yizheng Chen},
  title       = {Vulnerability Detection with Code Language Models: How Far are We?},
  booktitle   = {Proceedings of the 2025 IEEE/ACM 47th International Conference on Software Engineering (ICSE 2025)},
  pages       = {1729--1741},
  publisher   = {IEEE Computer Society},
  year        = {2025},
  doi         = {10.1109/ICSE55347.2025.00038}
}

@inproceedings{ethayarajh2019contextual,
  author      = {Kawin Ethayarajh},
  title       = {How Contextual are Contextualized Word Representations? Comparing the Geometry of {BERT}, {ELMo}, and {GPT-2} Embeddings},
  booktitle   = {Proceedings of the 2019 Conference on Empirical Methods in Natural Language Processing and the 9th International Joint Conference on Natural Language Processing (EMNLP-IJCNLP)},
  pages       = {55--65},
  publisher   = {Association for Computational Linguistics},
  year        = {2019},
  doi         = {10.18653/v1/D19-1006}
}

@article{fawcett2006introduction,
  author      = {Tom Fawcett},
  title       = {An Introduction to {ROC} Analysis},
  journal     = {Pattern Recognition Letters},
  volume      = {27},
  number      = {8},
  pages       = {861--874},
  publisher   = {Elsevier},
  year        = {2006},
  doi         = {10.1016/j.patrec.2005.10.010}
}

@inproceedings{fu2022linevul,
  author      = {Michael Fu and
                  Chakkrit Tantithamthavorn},
  title       = {{LineVul}: A Transformer-Based Line-Level Vulnerability Prediction},
  booktitle   = {Proceedings of the 19th International Conference on Mining Software Repositories (MSR '22)},
  pages       = {608--620},
  publisher   = {Association for Computing Machinery},
  year        = {2022},
  doi         = {10.1145/3524842.3528452}
}

@article{geirhos2020shortcut,
  author      = {Robert Geirhos and
                  J{\"o}rn-Henrik Jacobsen and
                  Claudio Michaelis and
                  Richard Zemel and
                  Wieland Brendel and
                  Matthias Bethge and
                  Felix A. Wichmann},
  title       = {Shortcut Learning in Deep Neural Networks},
  journal     = {Nature Machine Intelligence},
  volume      = {2},
  number      = {11},
  pages       = {665--673},
  publisher   = {Nature Publishing Group},
  year        = {2020},
  doi         = {10.1038/s42256-020-00257-z}
}

@inproceedings{gunel2020supervised,
  author      = {Beliz Gunel and
                  Jingfei Du and
                  Alexis Conneau and
                  Veselin Stoyanov},
  title       = {Supervised Contrastive Learning for Pre-trained Language Model Fine-tuning},
  booktitle   = {9th International Conference on Learning Representations (ICLR 2021)},
  publisher   = {OpenReview.net},
  year        = {2021},
  url         = {https://openreview.net/forum?id=cuBGzpB7Sy}
}

@inproceedings{guo2017calibration,
  author      = {Chuan Guo and
                  Geoff Pleiss and
                  Yu Sun and
                  Kilian Q. Weinberger},
  title       = {On Calibration of Modern Neural Networks},
  booktitle   = {Proceedings of the 34th International Conference on Machine Learning (ICML 2017)},
  series      = {Proceedings of Machine Learning Research},
  volume      = {70},
  pages       = {1321--1330},
  publisher   = {PMLR},
  year        = {2017},
  url         = {http://proceedings.mlr.press/v70/guo17a.html}
}

@inproceedings{hendrycks2017baseline,
  author      = {Dan Hendrycks and
                  Kevin Gimpel},
  title       = {A Baseline for Detecting Misclassified and Out-of-Distribution Examples in Neural Networks},
  booktitle   = {5th International Conference on Learning Representations (ICLR 2017)},
  publisher   = {OpenReview.net},
  year        = {2017},
  url         = {https://openreview.net/forum?id=Hkg4TI9xl}
}

@inproceedings{hu2021lora,
  author      = {Edward J. Hu and
                  Yelong Shen and
                  Phillip Wallis and
                  Zeyuan Allen-Zhu and
                  Yuanzhi Li and
                  Shean Wang and
                  Lu Wang and
                  Weizhu Chen},
  title       = {{LoRA}: Low-Rank Adaptation of Large Language Models},
  booktitle   = {The Tenth International Conference on Learning Representations (ICLR 2022)},
  publisher   = {OpenReview.net},
  year        = {2022},
  url         = {https://openreview.net/forum?id=nZeVKeeFYf9}
}

@inproceedings{huang2021whiteningbert,
  author      = {Junjie Huang and
                  Duyu Tang and
                  Wanjun Zhong and
                  Shuai Lu and
                  Linjun Shou and
                  Ming Gong and
                  Daxin Jiang and
                  Nan Duan},
  title       = {{WhiteningBERT}: An Easy Unsupervised Sentence Embedding Approach},
  booktitle   = {Findings of the Association for Computational Linguistics: EMNLP 2021},
  pages       = {238--244},
  publisher   = {Association for Computational Linguistics},
  year        = {2021},
  doi         = {10.18653/v1/2021.findings-emnlp.23}
}

@inproceedings{huang2025risk,
  author      = {Yuheng Huang and
                  Lei Ma and
                  Keizaburo Nishikino and
                  Takumi Akazaki},
  title       = {Risk Assessment Framework for Code {LLMs} via Leveraging Internal States},
  booktitle   = {Proceedings of the 33rd ACM International Conference on the Foundations of Software Engineering (FSE Companion '25)},
  pages       = {432--443},
  publisher   = {Association for Computing Machinery},
  year        = {2025},
  doi         = {10.1145/3696630.3728566}
}

@article{hui2024qwen25coder,
  author      = {Binyuan Hui and
                  Jian Yang and
                  Zeyu Cui and
                  Jiaxi Yang and
                  Dayiheng Liu and
                  Lei Zhang and
                  Tianyu Liu and
                  Jiajun Zhang and
                  Bowen Yu and
                  Kai Dang and
                  An Yang and
                  Rui Men and
                  Junyang Lin},
  title       = {{Qwen2.5-Coder} Technical Report},
  journal     = {arXiv preprint arXiv:2409.12186},
  year        = {2024},
  url         = {https://arxiv.org/abs/2409.12186}
}

@article{jiang2025investigatingllmcvd,
  author      = {Xuefeng Jiang and
                  Lvhua Wu and
                  Sheng Sun and
                  Jia Li and
                  Jingjing Xue and
                  Yuwei Wang and
                  Tingting Wu and
                  Min Liu},
  title       = {Investigating Large Language Models for Code Vulnerability Detection: An Experimental Study},
  journal     = {arXiv preprint arXiv:2412.18260},
  year        = {2024},
  url         = {https://arxiv.org/abs/2412.18260}
}

@inproceedings{jimenez2019importance,
  author      = {Matthieu Jimenez and
                  Renaud Rwemalika and
                  Mike Papadakis and
                  Federica Sarro and
                  Yves Le Traon and
                  Mark Harman},
  title       = {The Importance of Accounting for Real-World Labelling When Predicting Software Vulnerabilities},
  booktitle   = {Proceedings of the 2019 27th ACM Joint Meeting on European Software Engineering Conference and Symposium on the Foundations of Software Engineering (ESEC/FSE 2019)},
  pages       = {695--705},
  publisher   = {Association for Computing Machinery},
  year        = {2019},
  doi         = {10.1145/3338906.3338941}
}

@inproceedings{khosla2020supcon,
  author      = {Prannay Khosla and
                  Piotr Teterwak and
                  Chen Wang and
                  Aaron Sarna and
                  Yonglong Tian and
                  Phillip Isola and
                  Aaron Maschinot and
                  Ce Liu and
                  Dilip Krishnan},
  title       = {Supervised Contrastive Learning},
  booktitle   = {Advances in Neural Information Processing Systems (NeurIPS 2020)},
  volume      = {33},
  pages       = {18661--18673},
  publisher   = {Curran Associates, Inc.},
  year        = {2020},
  url         = {https://proceedings.neurips.cc/paper/2020/hash/d89a66c7c80a29b1bdbab0f2a1a94af8-Abstract.html}
}

@inproceedings{kumar2022finetuningdistort,
  author      = {Ananya Kumar and
                  Aditi Raghunathan and
                  Robbie Matthew Jones and
                  Tengyu Ma and
                  Percy Liang},
  title       = {Fine-Tuning Can Distort Pretrained Features and Underperform Out-of-Distribution},
  booktitle   = {The Tenth International Conference on Learning Representations (ICLR 2022)},
  publisher   = {OpenReview.net},
  year        = {2022},
  url         = {https://openreview.net/forum?id=RYSpflr_adC}
}

@inproceedings{lbath2026toward,
  author      = {Amine Lbath},
  title       = {Toward Scalable Automated Repository-Level Datasets for Software Vulnerability Detection},
  booktitle   = {Proceedings of the 34th ACM International Conference on the Foundations of Software Engineering (FSE Companion '26)},
  pages       = {23--24},
  publisher   = {Association for Computing Machinery},
  year        = {2026},
  doi         = {10.1145/3803437.3804870}
}

@inproceedings{lee2018mahalanobis,
  author      = {Kimin Lee and
                  Kibok Lee and
                  Honglak Lee and
                  Jinwoo Shin},
  title       = {A Simple Unified Framework for Detecting Out-of-Distribution Samples and Adversarial Attacks},
  booktitle   = {Advances in Neural Information Processing Systems (NeurIPS 2018)},
  volume      = {31},
  pages       = {7167--7177},
  publisher   = {Curran Associates, Inc.},
  year        = {2018},
  url         = {https://proceedings.neurips.cc/paper/2018/hash/abdeb6f575ac5c6676b747bca8d09cc2-Abstract.html}
}

@article{lee2019freezing,
  author      = {Jaejun Lee and
                  Raphael Tang and
                  Jimmy Lin},
  title       = {What Would {ELSA} Do? Freezing Layers During Transformer Fine-Tuning},
  journal     = {arXiv preprint arXiv:1911.03090},
  year        = {2019},
  url         = {https://arxiv.org/abs/1911.03090}
}

@inproceedings{li2025outofdistribution,
  author      = {Yikun Li and
                  Ngoc Tan Bui and
                  Ting Zhang and
                  Chengran Yang and
                  Xin Zhou and
                  Martin Weyssow and
                  Jinfeng Jiang and
                  Junkai Chen and
                  Huihui Huang and
                  Huu Hung Nguyen and
                  Chiok Yew Ho and
                  Jie Tan and
                  Ruiyin Li and
                  Yide Yin and
                  Han Wei Ang and
                  Frank Liauw and
                  Eng Lieh Ouh and
                  Lwin Khin Shar and
                  David Lo},
  title       = {Out of Distribution, Out of Luck: How Well Can {LLMs} Trained on Vulnerability Datasets Detect Top 25 {CWE} Weaknesses?},
  booktitle   = {Proceedings of the 2026 IEEE/ACM 48th International Conference on Software Engineering (ICSE '26)},
  pages       = {2096--2108},
  publisher   = {Association for Computing Machinery},
  year        = {2026},
  doi         = {10.1145/3744916.3787830}
}

@article{mclvd,
  author      = {Yi Cao and
                  Xiaolin Ju and
                  Xiang Chen and
                  Lina Gong},
  title       = {{MCL-VD}: Multi-modal Contrastive Learning with {LoRA}-Enhanced {GraphCodeBERT} for Effective Vulnerability Detection},
  journal     = {Automated Software Engineering},
  volume      = {32},
  number      = {2},
  pages       = {67},
  publisher   = {Springer},
  year        = {2025},
  doi         = {10.1007/s10515-025-00543-3}
}

@inproceedings{nashid2025multihunk,
  author      = {Noor Nashid and
                  Daniel Ding and
                  Keheliya Gallaba and
                  Ahmed E. Hassan and
                  Ali Mesbah},
  title       = {Characterizing Multi-Hunk Patches: Divergence, Proximity, and {LLM} Repair Challenges},
  booktitle   = {Proceedings of the 40th IEEE/ACM International Conference on Automated Software Engineering (ASE '25)},
  pages       = {1629--1641},
  publisher   = {IEEE},
  year        = {2025},
  doi         = {10.1109/ASE63991.2025.00137}
}

@misc{nvdapi,
  author      = {{{National Institute of Standards and Technology}}},
  title       = {{NVD} Vulnerability {APIs}},
  note        = {Accessed: 2026-02-15},
  year        = {2024},
  url         = {https://nvd.nist.gov/developers/vulnerabilities}
}

@inproceedings{ovadia2019trust,
  author      = {Yaniv Ovadia and
                  Emily Fertig and
                  Jie Ren and
                  Zachary Nado and
                  D. Sculley and
                  Sebastian Nowozin and
                  Joshua Dillon and
                  Balaji Lakshminarayanan and
                  Jasper Snoek},
  title       = {Can You Trust Your Model's Uncertainty? Evaluating Predictive Uncertainty Under Dataset Shift},
  booktitle   = {Advances in Neural Information Processing Systems (NeurIPS 2019)},
  volume      = {32},
  pages       = {13991--14002},
  publisher   = {Curran Associates, Inc.},
  year        = {2019},
  url         = {https://proceedings.neurips.cc/paper/2019/hash/8558cb408c1d76621371888657d2eb1d-Abstract.html}
}

@inproceedings{pan2026mitigating,
  author      = {Shengyi Pan and
                  Zhongxin Liu and
                  Jiayuan Zhou and
                  Xing Hu and
                  Xin Xia and
                  Shanping Li},
  title       = {Mitigating Implicit Inconsistencies in Patch Porting},
  booktitle   = {Proceedings of the 34th ACM International Conference on the Foundations of Software Engineering (FSE Companion '26)},
  pages       = {762--772},
  publisher   = {Association for Computing Machinery},
  year        = {2026},
  doi         = {10.1145/3803437.3805248}
}

@article{papernot2018deepknn,
  author      = {Nicolas Papernot and
                  Patrick McDaniel},
  title       = {Deep k-Nearest Neighbors: Towards Confident, Interpretable and Robust Deep Learning},
  journal     = {arXiv preprint arXiv:1803.04765},
  year        = {2018},
  url         = {https://arxiv.org/abs/1803.04765}
}

@misc{rahman2023slicing,
  author      = {Md Mahbubur Rahman and
                  Md Anwar Hossen},
  title       = {Vulnerability Detection Using Program Slicing and Transformer-Based Model},
  howpublished = {Preprint},
  year        = {2023},
  url         = {https://mahbubcseju.com/uploads/slice_level_vul_det.pdf}
}

@inproceedings{ran2025efficient,
  author      = {Dezhi Ran and
                  Lin Li and
                  Liuchuan Zhu and
                  Yuan Cao and
                  Landelong Zhao and
                  Xin Tan and
                  Guangtai Liang and
                  Qianxiang Wang and
                  Tao Xie},
  title       = {Efficient and Robust Security-Patch Localization for Disclosed {OSS} Vulnerabilities with Fine-Tuned {LLMs} in an Industrial Setting},
  booktitle   = {Proceedings of the 33rd ACM International Conference on the Foundations of Software Engineering (FSE Companion '25)},
  pages       = {262--273},
  publisher   = {Association for Computing Machinery},
  year        = {2025},
  doi         = {10.1145/3696630.3728551}
}

@inproceedings{robinson2021contrastive,
  author      = {Joshua David Robinson and
                  Ching-Yao Chuang and
                  Suvrit Sra and
                  Stefanie Jegelka},
  title       = {Contrastive Learning with Hard Negative Samples},
  booktitle   = {9th International Conference on Learning Representations (ICLR 2021)},
  publisher   = {OpenReview.net},
  year        = {2021},
  url         = {https://openreview.net/forum?id=CR1XOQ0UTh-}
}

@article{sclcvd,
  author      = {Rongcun Wang and
                  Senlei Xu and
                  Yuan Tian and
                  Xingyu Ji and
                  Xiaobing Sun and
                  Shujuan Jiang},
  title       = {{SCL-CVD}: Supervised Contrastive Learning for Code Vulnerability Detection via {GraphCodeBERT}},
  journal     = {Computers \& Security},
  volume      = {145},
  pages       = {103994},
  publisher   = {Elsevier},
  year        = {2024},
  doi         = {10.1016/j.cose.2024.103994}
}

@article{sheng2025llmssoftwaresecuritysurvey,
  author      = {Ze Sheng and
                  Zhicheng Chen and
                  Shuning Gu and
                  Heqing Huang and
                  Guofei Gu and
                  Jeff Huang},
  title       = {{LLMs} in Software Security: A Survey of Vulnerability Detection Techniques and Insights},
  journal     = {ACM Computing Surveys},
  volume      = {58},
  number      = {3},
  pages       = {65:1--65:39},
  publisher   = {Association for Computing Machinery},
  year        = {2025},
  doi         = {10.1145/3769082}
}

@article{shestov2024finetuning,
  author      = {Aleksei Shestov and
                  Rodion Levichev and
                  Ravil Mussabayev and
                  Evgeny Maslov and
                  Pavel Zadorozhny and
                  Anton Cheshkov and
                  Rustam Mussabayev and
                  Alymzhan Toleu and
                  Gulmira Tolegen and
                  Alexander Krassovitskiy},
  title       = {Finetuning Large Language Models for Vulnerability Detection},
  journal     = {IEEE Access},
  volume      = {13},
  pages       = {38889--38900},
  publisher   = {IEEE},
  year        = {2025},
  doi         = {10.1109/ACCESS.2025.3546700}
}

@inproceedings{shimmi2024vulsim,
  author      = {Samiha Shimmi and
                  Ashiqur Rahman and
                  Mohan Gadde and
                  Hamed Okhravi and
                  Mona Rahimi},
  title       = {{VulSim}: Leveraging Similarity of Multi-Dimensional Neighbor Embeddings for Vulnerability Detection},
  booktitle   = {33rd USENIX Security Symposium (USENIX Security 24)},
  pages       = {2661--2678},
  publisher   = {USENIX Association},
  year        = {2024},
  url         = {https://www.usenix.org/conference/usenixsecurity24/presentation/shimmi}
}

@inproceedings{steenhoek2023empirical,
  author      = {Benjamin Steenhoek and
                  Md Mahbubur Rahman and
                  Richard Jiles and
                  Wei Le},
  title       = {An Empirical Study of Deep Learning Models for Vulnerability Detection},
  booktitle   = {Proceedings of the 2023 IEEE/ACM 45th International Conference on Software Engineering (ICSE '23)},
  pages       = {2237--2248},
  publisher   = {IEEE Computer Society},
  year        = {2023},
  doi         = {10.1109/ICSE48619.2023.00188}
}

@article{su2021whitening,
  author      = {Jianlin Su and
                  Jiarun Cao and
                  Weijie Liu and
                  Yangyiwen Ou},
  title       = {Whitening Sentence Representations for Better Semantics and Faster Retrieval},
  journal     = {arXiv preprint arXiv:2103.15316},
  year        = {2021},
  url         = {https://arxiv.org/abs/2103.15316}
}

@inproceedings{sun2022knnood,
  author      = {Yiyou Sun and
                  Yifei Ming and
                  Xiaojin Zhu and
                  Yixuan Li},
  title       = {Out-of-Distribution Detection with Deep Nearest Neighbors},
  booktitle   = {Proceedings of the 39th International Conference on Machine Learning (ICML 2022)},
  series      = {Proceedings of Machine Learning Research},
  volume      = {162},
  pages       = {20656--20679},
  publisher   = {PMLR},
  year        = {2022},
  url         = {https://proceedings.mlr.press/v162/sun22d.html}
}

@inproceedings{tang2026benchmarking,
  author      = {Zhenghao Tang and
                  Zhenyu Zang and
                  Kaifeng Huang},
  title       = {Benchmarking {MCP} Vulnerability Detection from a Paired Evaluation Perspective},
  booktitle   = {Proceedings of the 34th ACM International Conference on the Foundations of Software Engineering (FSE Companion '26)},
  pages       = {1578--1583},
  publisher   = {Association for Computing Machinery},
  year        = {2026},
  doi         = {10.1145/3803437.3805534}
}

@inproceedings{tenney2019bert,
  author      = {Ian Tenney and
                  Dipanjan Das and
                  Ellie Pavlick},
  title       = {{BERT} Rediscovers the Classical {NLP} Pipeline},
  booktitle   = {Proceedings of the 57th Annual Meeting of the Association for Computational Linguistics (ACL 2019)},
  pages       = {4593--4601},
  publisher   = {Association for Computational Linguistics},
  year        = {2019},
  doi         = {10.18653/v1/P19-1452}
}

@article{vulgate,
  author      = {Rijha Safdar and
                  Danyail Mateen and
                  Syed Taha Ali and
                  Umer Ashfaq and
                  Wajahat Hussain},
  title       = {Data and Context Matter: Towards Generalizing {AI}-based Software Vulnerability Detection},
  journal     = {International Journal of Information Security},
  publisher   = {Springer},
  year        = {2026},
  doi         = {10.1007/s10207-026-01282-6}
}

@inproceedings{wang2025rag,
  author      = {Chaozheng Wang and
                  Zezhou Yang and
                  Shuzheng Gao and
                  Cuiyun Gao and
                  Ting Peng and
                  Hailiang Huang and
                  Yuetang Deng and
                  Michael R. Lyu},
  title       = {{RAG} or Fine-tuning? A Comparative Study on {LCMs}-based Code Completion in Industry},
  booktitle   = {Proceedings of the 33rd ACM International Conference on the Foundations of Software Engineering (FSE Companion '25)},
  pages       = {93--104},
  publisher   = {Association for Computing Machinery},
  year        = {2025},
  doi         = {10.1145/3696630.3728535}
}

@inproceedings{wu2024condefects,
  author      = {Yonghao Wu and
                  Zheng Li and
                  Jie M. Zhang and
                  Yong Liu},
  title       = {{ConDefects}: A Complementary Dataset to Address the Data Leakage Concern for {LLM}-Based Fault Localization and Program Repair},
  booktitle   = {Companion Proceedings of the 32nd ACM International Conference on the Foundations of Software Engineering (FSE Companion '24)},
  pages       = {642--646},
  publisher   = {Association for Computing Machinery},
  year        = {2024},
  doi         = {10.1145/3663529.3663815}
}

@inproceedings{xiao2020mvp,
  author      = {Yang Xiao and
                  Bihuan Chen and
                  Chendong Yu and
                  Zhengzi Xu and
                  Zimu Yuan and
                  Feng Li and
                  Binghong Liu and
                  Yang Liu and
                  Wei Huo and
                  Wei Zou and
                  Wenchang Shi},
  title       = {{MVP}: Detecting Vulnerabilities using {Patch-Enhanced} Vulnerability Signatures},
  booktitle   = {29th USENIX Security Symposium (USENIX Security 20)},
  pages       = {1165--1182},
  publisher   = {USENIX Association},
  year        = {2020},
  url         = {https://www.usenix.org/conference/usenixsecurity20/presentation/xiao}
}

@inproceedings{zhang2024lmpace,
  author      = {Dylan Zhang and
                  Xuchao Zhang and
                  Chetan Bansal and
                  Pedro Henrique B. Las-Casas and
                  Rodrigo Fonseca and
                  Saravan Rajmohan},
  title       = {{LM-PACE}: Confidence Estimation by Large Language Models for Effective Root Causing of Cloud Incidents},
  booktitle   = {Companion Proceedings of the 32nd ACM International Conference on the Foundations of Software Engineering (FSE Companion '24)},
  pages       = {388--398},
  publisher   = {Association for Computing Machinery},
  year        = {2024},
  doi         = {10.1145/3663529.3663858}
}

\end{document}